\RequirePackage{ifpdf}
\ifpdf
\documentclass[pdftex,epj,final]{svjour}
\else
\documentclass[epj,final]{svjour}
\fi
\usepackage{graphicx}
\ifpdf
\usepackage{epstopdf}
\fi
\usepackage{booktabs}

\begin{document}
%

\title{Saturation spectra of low lying states of Nitrogen:\\ reconciling experiment with theory}

\author{Thomas Carette\inst{1}, Messaoud Nemouchi\inst{2} 
, Per J\"{o}nsson\inst{3} and Michel Godefroid\inst{1}\thanks{\email{mrgodef@ulb.ac.be}}
%
}                     
%
%
\institute{Service de Chimie quantique et Photophysique, CP160/09, \\ 
Universit\'e Libre de Bruxelles, 
Avenue F.D. Roosevelt 50, B~1050 Brussels, Belgium 
\and Laboratoire d'\'Electronique Quantique, Facult\'e de Physique, USTHB, 
BP32, El-Alia, Algiers, Algeria 
\and Center for Technology Studies, Malm\"{o} University, 205-06 Malm\"{o}, Sweden}
\date{\today}
%
\abstract{
The hyperfine constants of the levels $2p^{2}(^{3}$P$)3s~^{4}$P$_{J}$, $2p^{2}(^{3}$P$)3p~^{4}$P$^{o}_{J}$ and $2p^{2}(^{3}$P$)3p~^{4}$D$^{o}_{J}$, deduced by  Jennerich et al.~[Eur. Phys. J. D {\bf 40}, 81 (2006)] from the observed hyperfine structures of the transitions $2p^{2}(^{3}$P$)3s~^{4}$P$_J \rightarrow 2p^{2}(^{3}$P$)3p~^{4}$P$^{o}_{J'}$ and $2p^{2}(^{3}$P$)3s~^{4}$P$_J \rightarrow  2p^{2}(^{3}$P$)3p~^{4}$D$^{o}_{J'}$ recorded by saturation spectroscopy in the near-infrared, 
 strongly disagree with the \emph{ab initio} values of J\"onsson et al.~[J.~Phys.~B: At. Mol. Opt. Phys. {\bf 43},115006 (2010)]. 
We propose a new interpretation of the recorded weak spectral lines. If the latter are indeed reinterpreted as crossover signals, a new set of experimental hyperfine constants is deduced, in very good agreement with the \emph{ab initio} predictions.
%
%
} 
\authorrunning{Carette {\em et al.}}
\titlerunning{Saturation spectra of low lying states of Nitrogen: reconciling experiment with theory}

\maketitle

\section{Introduction}
In 1943, Holmes \cite{Hol:43a} measured the isotope shifts (IS) of the $2p^{2}(^{3}$P$)3s~^{4}$P$_J \rightarrow  2p^{2}(^{3}$P$)3p~^{4}$P$^{o}_{J'}$, $2p^{2}(^{3}$P$)3s~^{2}$P$_J \rightarrow  2p^{2}(^{3}$P$)3p~^{2}$P$^{o}_{J'}$ and $2p^{2}(^{3}$P$)3s~^{4}$P$_J \rightarrow  2p^{2}(^{3}$P$)3p~^{4}$S$^{o}_{J'}$ transitions for the $^{15}$N$- ^{14}$N isotopic pair. He observed a surprising variation of the IS from one multiplet component to another of the same transition.
Cangiano et al.~\cite{Canetal:94a} later confirmed this effect by measuring the \mbox{hyperfine} structure constants and isotope shifts of the 
$2p^{2}(^{3}$P$)3s~^{4}$P$_{J} $ $\rightarrow $ $ 2p^{2}(^{3}$P$)3p~^{4}$P$^{o}_{J'}$ transitions using an external cavity diode laser and Doppler-free techniques. More recently, the hyperfine structures of these near-infrared transitions have been remeasured by Jennerich et al. \cite{Jenetal:06a}, also using saturation absorption spectroscopy but improving the spectral resolution. 
In the same work, the authors completed this study by investigating the structure of  $2p^{2}(^{3}$P$)3s~^{4}$P$_{J}$ $ \rightarrow$ $2p^{2}(^{3}$P$)3p~^{4}$D$^{o}_{J'}$ transitions around $870$~nm. Values of the hyperfine structure coupling constants of all the upper and lower multiplets were obtained for both isotopes. Isotope shifts of three transitions in each multiplet were also measured and the significant $J$-dependence of the shifts was confirmed.  The authors appealed for further theoretical investigation to confirm the observations.  

In response to this,
J\"onsson et al.~\cite{Jonetal:10a} calculated the electronic hyperfine factors using elaborate correlation models. The resulting  {\em ab initio} hyperfine constants disagree completely with the experimental parameters obtained by fitting the observed hyperfine spectra~\cite{Jenetal:06a}.
This disagreement calls for a reinterpretation of the experimental spectral lines.

The saturated-absorption spectroscopy is a Doppler-free method which measures the absorption of a probe beam in an atomic vapor cell saturated by a counter-propagating pump beam. The absorption spectrum of the probe beam featured several Lamb-dips with a width of the order of the natural width.  When the Doppler-broade\-ned line spreads on several transitions, as for instance in hyperfine spectra, crossover signals often appear when the pump and probe laser beams frequency corresponds to the average of the frequencies of two hyperfine transitions~\cite{Dem:08a}. A crossover signal might then show up in a spectrum between hyperfine lines sharing either the lower level or the upper level (involving three levels), or none of them (involving four levels)~\cite{Hanetal:71a,Andetal:78a}. In the former case, if the common level is the lower one, the two beams propagating through the atomic vapor both contribute in reducing its population. The probe beam absorption then weakens, like the absorption hyperfine lines. This corresponds to a positive intensity crossover. If the common level is the upper one, the probe beam absorption signal may either increase or decrease~\cite{Hanetal:71a,Andetal:78a,TatWal:99a}. 
If the probe beam absorption spectrum is resolved and strong enough, crossover signals might be helpful in identifying unambiguously the hyperfine lines~\cite{Krietal:09a}. 
More frequently, their presence complicates the spectral analysis due to possible overlaps with the hyperfine components themselves. Moreover, the theory of crossover intensities is rather complex. A sign inversion of the crossover intensities has been observed in the hyperfine spectrum of the sodium D1 line with a change of the vapor temperature~\cite{GurSha:99a}. 
Saturation effects, optical pumping~\cite{Sanetal:09a,Imetal:01a,Papetal:80a,Rinetal:80a},
radiation pressure~\cite{GriMly:89a}, pump and probe beam-polarizations~\cite{Schetal:94b}, may also affect the intensities of hyperfine lines  and crossover signals. Recent progress has been achieved in saturated absorption spectroscopy to eliminate crossovers in hyperfine spectra. The hyperfine structure spectrum of the rubidium D2 line has been so measured~\cite{Saretal:08a} using a vapor nano-cell. The same spectrum, measured in saturated absorption spectroscopy, with copropagating pump and probe  laser beams of the same intensity, is  also free of crossovers~\cite{BanNat:03a}.
 
While the strong hyperfine lines are relatively easy to identify, the weak components are usually not. The existence of crossover resonances as a consequence of the saturated absorption technique was recognized in only two transitions studied by Jennerich et al.~\cite{Jenetal:06a}. In the present work, we completely revisit their saturation spectra, calling their line assignments in question. Starting from the fact that the strong mismatch between observation and theory only concerns the weak hyperfine lines (see section~\ref{sec:hfs_spectra}), we reinterpret most of them as crossover signals (see section~\ref{sec:cross}). The new set of hyperfine constants is consistent with the {\em ab initio} results of J\"onsson et al.~\cite{Jonetal:10a}.

\section{Hyperfine spectra simulations}
\label{sec:hfs_spectra}
\begin{table*}[th!]
\begin{center}
\caption{Comparison between the original experimental~\cite{Jenetal:06a} and the {\em ab initio}~\cite{Jonetal:10a} hyperfine constants with the new values determined from the present analysis. All values are in MHz.\label{AJ}}
\begin{tabular}{cccccccccc}
\toprule
 & \multicolumn{3}{c}{$^{15}$N} & \multicolumn{6}{c}{$^{14}$N} \\
 \cmidrule(r){2-4}  \cmidrule(l){5-10}
State & Exp.~\cite{Jenetal:06a} & Theory~\cite{Jonetal:10a} & This work &\multicolumn{2}{c}{Exp.~\cite{Jenetal:06a}} & \multicolumn{2}{c}{Theory~\cite{Jonetal:10a}} & \multicolumn{2}{c}{This work} \\
& $A$ & $A$ & $A$&$A$ & $B$ & $A$ & $B$ & $A$ & $B$\\
\midrule
$^4$P$_{1/2}$ &$\left\{\begin{array}{c}
                        +103.4(14)\\
                        -153.1(23)^{a}\\ 
                        \end{array} \right. $ & $-$140.56 & $-$153.1(23) &$\left\{\begin{array}{c}
                        -69.76(90) \\
                        +112.3(13)^{a}\\ 
                        \end{array} \right.$ & 0.0 & 100.21 & 0.0 & 112.3(13) &0.0  \\
$^4$P$_{3/2}$ & $-$47.93(48) & $-$87.62 & $-$95.86(96) & 35.52(44) & $-$0.98(48) & 62.46 & 4.10 & 68.33(69)$^b$ & 3.5(91) \\
$^4$P$_{5/2}$ & $-$90.71(71) & $-$175.12 & $-$181.4(15) & 64.76(42) & $-$3.9(10) & 124.84 & $-5.12$ & 129.52(84) & $-$7.8(20) \\
$^4$P$^o_{1/2}$ & 167.1(13) & 73.29 & 71.2(23)  &$-$133.2(22) & 0.0 & $-$52.25 & 0.0 & $-$50.78(17)$^b$ & 0.0   \\
$^4$P$^o_{3/2}$ & 70.0(12) & $-$71.60 & $-$66.1(23) & $-$48.56(74) & 8.69(87) & 51.04 & $-$2.95 & 46.2(15) & $-$2.7(17) \\
$^4$P$^o_{5/2}$ & 46.20(74) & $-$46.52 & $-$44.5(15) & $-$32.83(44) & 5.0(11) & 33.16 & 2.57 & 31.93(86) & 1.1(21) \\
$^4$D$^o_{1/2}$ &$\left\{\begin{array}{c}
                       +153.1(23)\\
                       -103.4(14)^{a}\\ 
                        \end{array} \right. $ & $-$104.02 & $-$103.4(14) & $\left\{\begin{array}{c}
                        -112.3(13)\\
                        +69.76(90)^{a}\\ 
                        \end{array} \right. $ & 0.0 & 74.15 & 0.0 & 69.76(90) & 0.0\\
$^4$D$^o_{3/2}$ & 92.4(17) & $-$44.49 & $-$43.7(28) & $-$64.41(79) & 10.46(88) & 31.71 & 0.30 & 30.3(15) & $-$0.9(17)  \\
$^4$D$^o_{5/2}$ & 41.5(14) & $-$51.57 & $-$49.2(22) & $-$28.19(62) & $-$0.2(15) & 36.76 & $-$1.69 & 36.6(11) & $-$4.1(25)  \\
$^4$D$^o_{7/2}$ & $-$9.35(55) & $-$78.04 & $-$77.4(11) & 6.31(72) & $-$12.6(13) & 55.63 & $-$6.44 & 55.2(11) & $-$9.9(26)  \\
\bottomrule
\multicolumn{10}{l}{%
\begin{minipage}{\textwidth}\footnotesize
\vspace{0.1cm}
$^a$ Second proposition of Jennerich et al.~\cite{Jenetal:06a} (see text).
\end{minipage}
}\\
\multicolumn{10}{l}{%
\begin{minipage}{\textwidth}\footnotesize
\vspace{0.1cm}
$^b$ Values taken from the constraint $A(^{15}$N$)/A(^{14}$N$)=-1.4028$.
\end{minipage}
}
\end{tabular}
\end{center}
\end{table*}

The hyperfine structure of a spectrum is caused by the interaction of the angular momentum of the electrons ($\textbf{J}$) and of the nucleus ($\textbf{I}$), forming the total atomic angular momentum $\textbf{F}= \textbf{I} + \textbf{J}$. Neglecting the higher order multipoles as well as the off-diagonal effects, the energy $W_{F}$ of a hyperfine level, characterized by the quantum number $F$ associated to $\textbf{F}$, is
\begin{equation}\label{ehyp}
W_{F}=W_{J}+ A\frac{C}{2}+ B\frac{3C(C+1)- 4 I(I+1)J(J+1)}{8I(2I-1)J(2J-1)}
\end{equation}
with $C=F(F+1)-I(I+1)-J(J+1)$. $W_{J}$ is the energy of the fine structure level $J$. $A$ and $B$ are the hyperfine constants that  describe respectively the magnetic dipole  and  electric quadrupole interactions.

Giving $A$ and $B$ in MHz, the frequency of a hyperfine transition between two levels ($JF$) and ($J'F'$) is: 
\begin{equation}\label{nuhf}
\nu=\nu_{0}+a'A'+b'B'-aA-bB
\end{equation}
where  the primed symbols stand for the upper level. 
$\nu_{0}$ is the frequency of the $J$-$J'$ transition.
The factors $a$ and $b$ ($a'$ and $b'$) are the coefficients that weight the hyperfine constants $A$ and $B$ ($A'$ and $B'$) in formula (\ref{ehyp}), i.e.
\begin{equation}
a=\frac{C}{2} \hspace*{0.5cm} ;  \hspace*{0.5cm} b=\frac{3C(C+1)-4I(I+1)J(J+1)}{8I(2I-1)J(2J-1)} \; .
\end{equation}
To be consistent with \cite{Jenetal:06a}, we simulate the spectra using Lorentzian line shapes with a 70~MHz width corresponding to the natural linewidth. 
The relative intensities of the hyperfine lines $I_{r}$ are deduced from the formula~\cite{Cow:81a}
\begin{equation}
\label{rel_int}
I_{r}=(2F+1)(2F'+1)
\left\{
\begin{array}{ccc}
J&I&F\\
F'&1&J'
\end{array}
\right\}^{2}\; .
\end{equation}

The total nuclear angular momentum  $I$ of the isotopes $^{15}$N and $^{14}$N are  equal to $1/2$ and $1$, respectively. The nuclear quadrupole moment 
$Q$ is non zero for the isotope $^{14}$N only,  $Q(^{14}$N)=+0.02001(10)~b~\cite{Sto:05a}. The nuclear magnetic moments of the isotopes are $\mu_{I}(^{15}$N$)=-0.28318884(5)$~nm and $\mu_{I}(^{14}$N)$=+0.40376100(6)$~nm~\cite{Sto:05a}. The expected ratio between the magnetic hyperfine constants characterizing a given $J$-level of the two isotopes should be
\begin{equation}
A_{J}(^{15}\mbox{N})/A_{J}(^{14}\mbox{N})= 
\frac{\mu_{I}(^{15}\mbox{N}) I(^{14}\mbox{N})}{\mu_{I}(^{14}\mbox{N}) I(^{15}\mbox{N})} =-1.4028\, .
\end{equation}

\begin{figure*}
\center
\resizebox{0.45\textwidth}{!}{\includegraphics{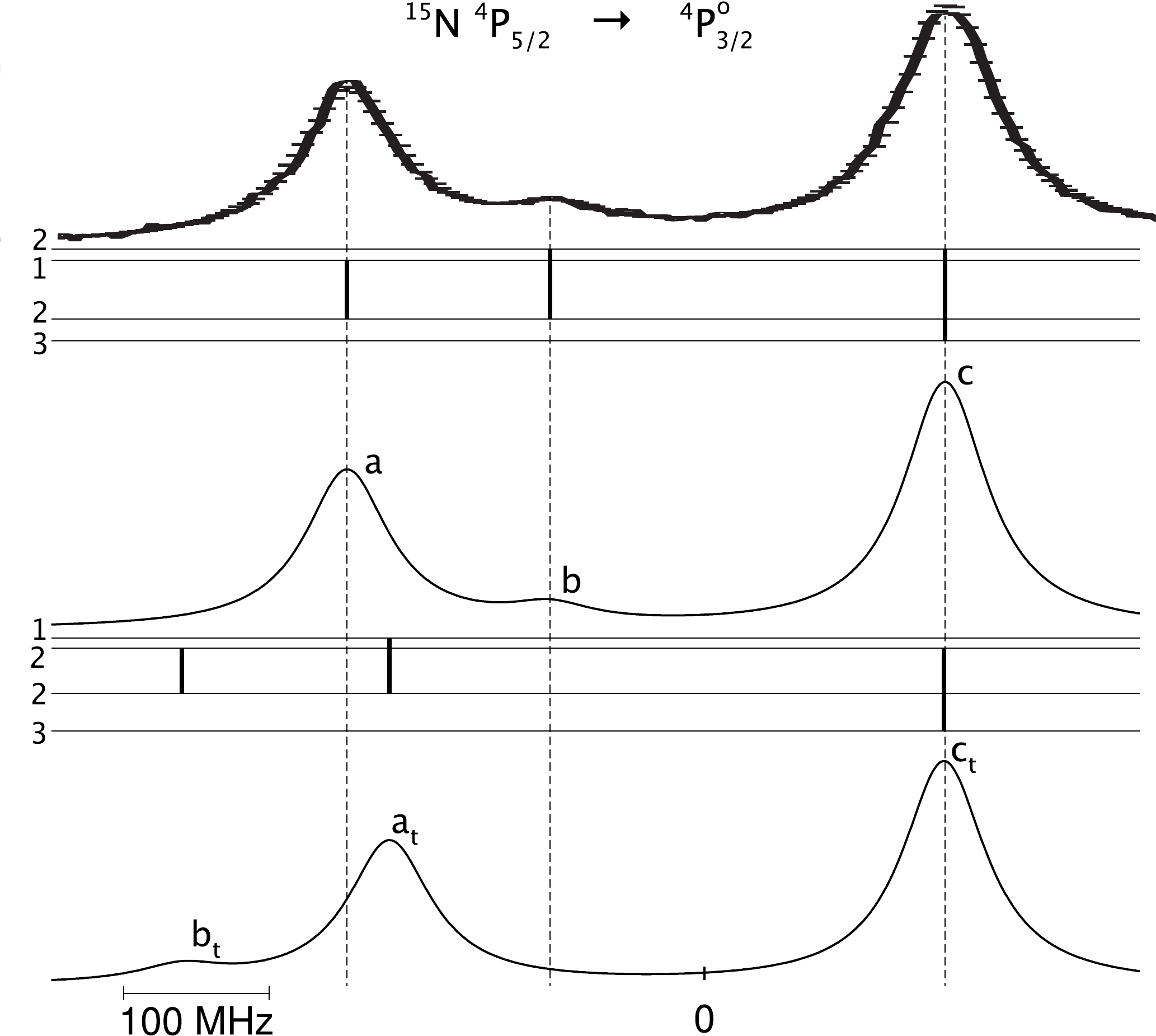}}
\resizebox{0.45\textwidth}{!}{\includegraphics{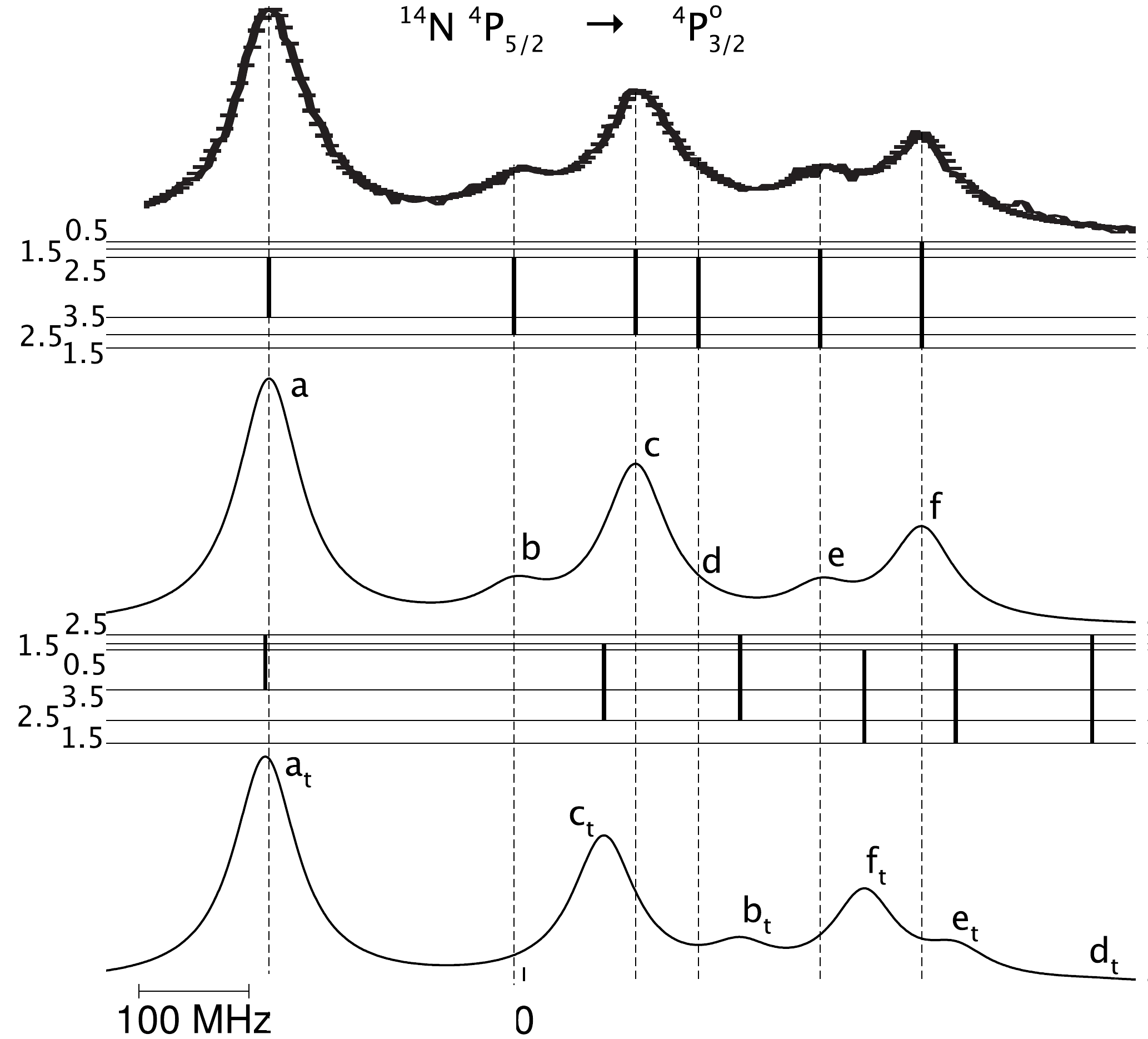}}
\caption{
Top : hyperfine spectra of the transition $^{4}$P$_{5/2}\rightarrow~^{4}$P$^{o}_{3/2}$ recorded by Jennerich et al.~\cite{Jenetal:06a} for both isotopes (digitized from the figures of the original article). Middle and bottom : level and transition  diagrams and corresponding simulated spectra, $\mathcal{S}$ and $\mathcal{S}_t$ (omitting the crossovers),  using respectively the experimental constants of Jennerich et al.~\cite{Jenetal:06a} and the {\em ab initio} theoretical constants of J\"onsson et al.~\cite{Jonetal:10a}. For the latter, a $t$-subscript is added to the line symbol. The used linewidth is 70 MHz. The center of gravity of both $\mathcal{S}$ and $\mathcal{S}_t$ is set to zero.
}\label{fig:Per_5/2_3/2}
\end{figure*}

\begin{figure*}
\center
\resizebox{0.45\textwidth}{!}{\includegraphics{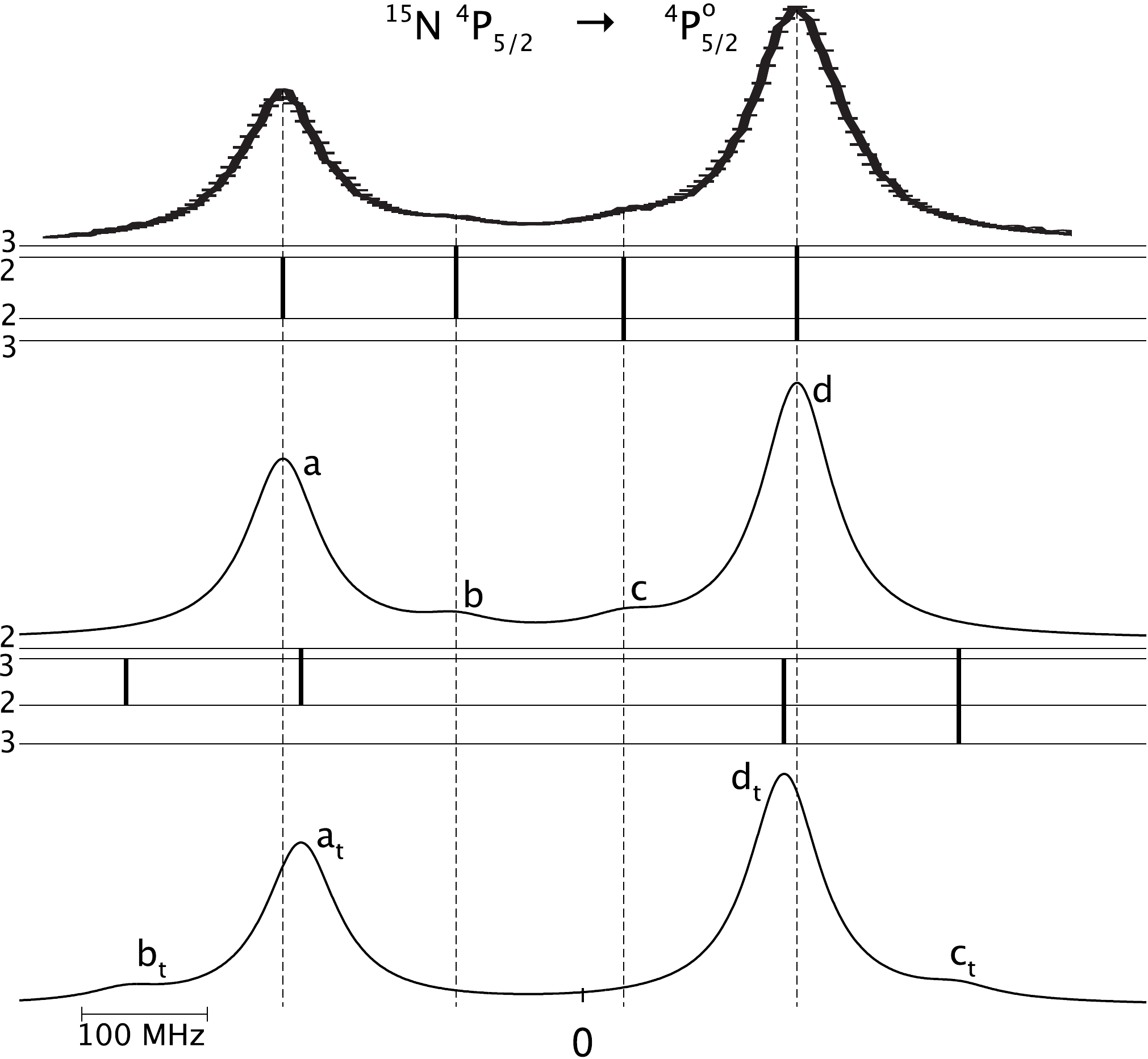}}
\resizebox{0.45\textwidth}{!}{\includegraphics{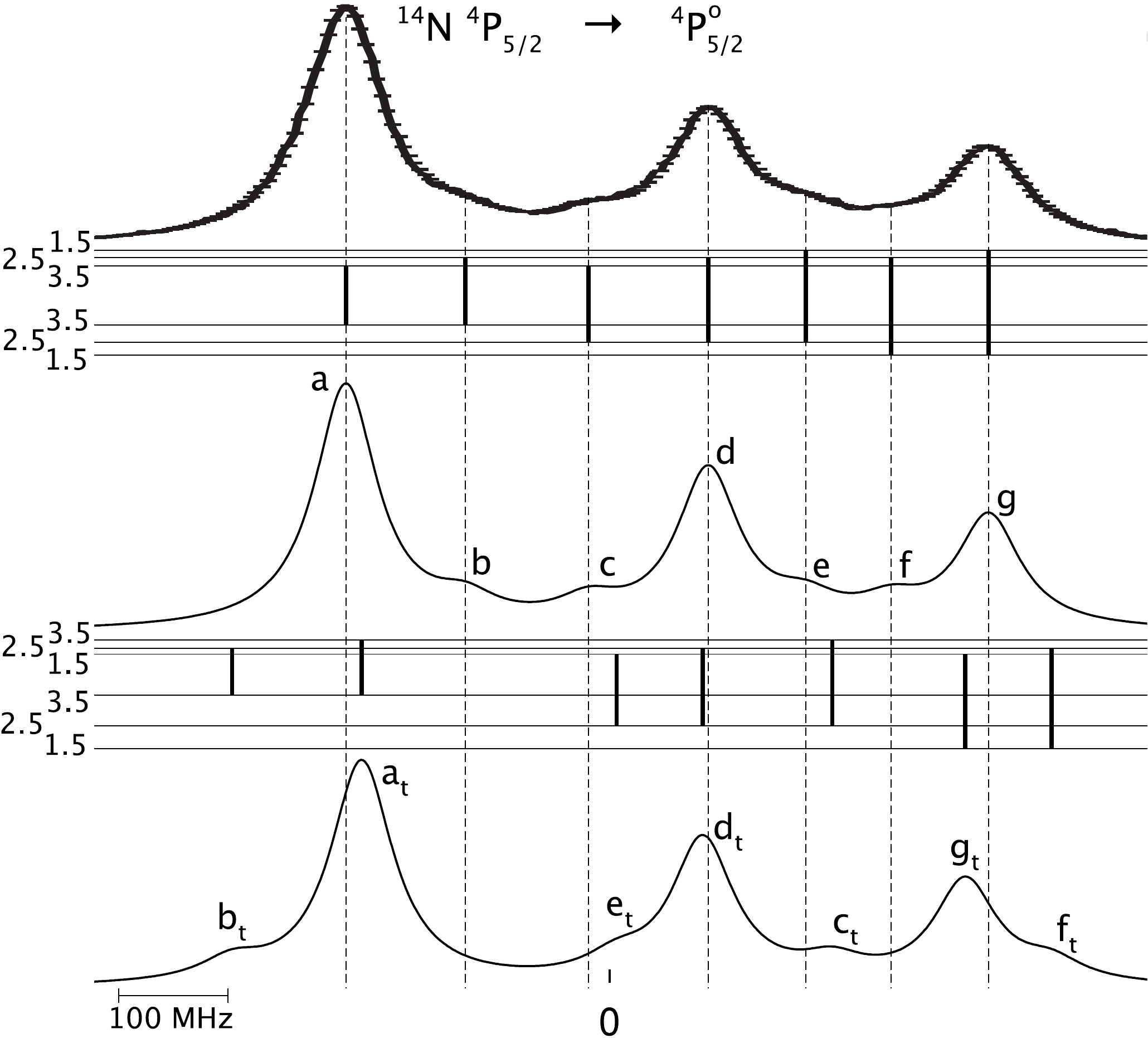}}
\caption{Top : hyperfine spectra of the transition $^{4}$P$_{5/2}\rightarrow~^{4}$P$^{o}_{5/2}$ recorded by Jennerich et al.~\cite{Jenetal:06a} for both isotopes (digitized from the figures of the original article). Middle and bottom : level and transition  diagrams and corresponding simulated spectra, $\mathcal{S}$ and $\mathcal{S}_t$ (omitting the crossovers), using respectively the experimental constants of Jennerich et al.~\cite{Jenetal:06a} and the {\em ab initio} theoretical constants of J\"onsson et al.~\cite{Jonetal:10a}. For the latter, a $t$-subscript is added to the line symbol. The used linewidth is 70 MHz.
The center of gravity of both $\mathcal{S}$ and $\mathcal{S}_t$ is set to zero.
}\label{fig:Per_5/2_5/2}
\end{figure*}

Table \ref{AJ} presents the {\em ab initio} hyperfine constants $A$ and $B$ of J\"onsson et al.~\cite{Jonetal:10a}, obtained from elaborate multiconfigurational Hartree-Fock calculations with relativistic corrections, together with the experimental ones of Jennerich et al.~\cite{Jenetal:06a}. As concluded in~\cite{Jonetal:10a}, the huge and systematic disagreement between observation and theory appeals for further investigations. We compare those two sets through their corresponding spectral simulations. Figures~\ref{fig:Per_5/2_3/2} and \ref{fig:Per_5/2_5/2} display the recorded and simulated spectra for transitions $3s \; ^{4}$P$_{5/2}$ 
$ \rightarrow$ $~3p \; ^{4}$P$^{o}_{3/2,5/2}$ of both isotopes $^{15}$N and $^{14}$N.
The upper spectra are the ones recorded by Jennerich~et al.~\cite{Jenetal:06a} (the dots that became short horizontal lines in the digitizing process of the original figures, correspond to the recorded data while the continuous line is the result of their fit).
The middle and bottom parts of the figures are the synthetic spectra calculated using the original experimental constants from~\cite{Jenetal:06a} and the \emph{ab initio} constants from~\cite{Jonetal:10a}, respectively, and are denoted hereafter $\mathcal{S}$ and $\mathcal{S}_t$. For the latter, we add a $t$-subscript to the letters $a, b, c, \dots$, characterizing the transitions. One observes that the huge disagreement between the theoretical and experimental hyperfine constants reported in Table~\ref{AJ} mostly concerns the weak line's positions, while the intense lines agree satisfactorily.

In Figure~\ref{fig:Per_5/2_3/2}, each simulation is accompanied by its corresponding level and transition diagrams specifying, for each hyperfine spectral line, the upper and lower $F$-values. From the diagram corresponding to the transition $^{4}$P$_{5/2}\rightarrow~^{4}$P$^{o}_{3/2}$ of $^{15}$N (left part of Figure~\ref{fig:Per_5/2_3/2}), one realizes that the lines $b$ and $c$ have a common upper level. It means that a crossover could appear at a frequency $(\nu_b+ \nu_c)/2$. The line $b$ of the simulated experimental spectra $\mathcal{S}$ could be reinterpreted as a crossover signal of lines $b_{t}$ and $c_{t}$ of the theoretical spectrum $\mathcal{S}_t$. Why the crossover signal $b={\mbox{co}} ( b_t,c_t)$ appears while the real line $b_t$ does not show up, is unclear.
Likewise, the experimental line $b$ of the $^{14}$N spectrum (right part of Figure~\ref{fig:Per_5/2_3/2}) can be reinterpreted as the crossover of $a_{t}$ and $b_{t}$ while $e$ could be the crossover of $c_{t}$ and $e_{t}$. These observations are the starting point of the present analysis. The same arguments apply to all low intensity lines of the experimental spectra of the transitions $^{4}$P$_{5/2}\rightarrow ~^{4}$P$^{o}_{5/2}$ (see Figure~\ref{fig:Per_5/2_5/2}). The hyperfine level diagrams of the transitions $^{4}$P$_{5/2}\rightarrow ~^{4}$D$^{o}_{3/2,5/2}$ differing from those of  $^{4}$P$_{5/2}\rightarrow ~^{4}$P$^{o}_{3/2,5/2}$ only by their upper level spacing, their spectra are alike and a similar reinterpretation of the experimental signals in terms of crossovers is possible. 

In the following section, we show that the {\em ab initio} hyperfine constants~\cite{Jonetal:10a} of the states $^{4}$P$^{o}_{3/2,5/2}$, $^{4}$D$^{o}_{3/2,5/2,7/2}$ and $^{4}$P$_{5/2}$ are compatible with the recorded spectra of Jennerich et al.~\cite{Jenetal:06a}, at the condition that we identify the low intensity lines as crossover signals. We also confirm the intense hyperfine line's identification.
We then discuss the hyperfine spectra corresponding to the transitions \mbox{$^{4}$P$_{3/2}\rightarrow ~^{4}$P$^{o}_{1/2}$}  and \mbox{$^{4}$P$_{1/2}\rightarrow ~^{4}$D$^{o}_{1/2}$}, which are analyzed somewhat differently. A new set of hyperfine constants is deduced and used to compare the unresolved experimental spectra  \mbox{$^{4}$P$_{1/2}\rightarrow~^{4}$P$^{o}_{3/2}$}, \mbox{$^{4}$P$_{3/2} \rightarrow ~^{4}$P$^o_{3/2}$} and \mbox{$^{4}$P$_{3/2}\rightarrow~^{4}$P$^{o}_{5/2}$} to the theoretical simulations.



\section{Interpretation of the weak lines in terms of crossovers}\label{sec:cross}

The procedure to deduce a new set of  hyperfine constants is based on the reinterpretation of the original spectra recorded by Jennerich et al.~\cite{Jenetal:06a}. The line
frequencies are recalculated from their original set of hyperfine constants using  equation~(\ref{nuhf}). 
The residuals (data minus fit) reported in~\cite{Jenetal:06a} 
  being small - about 4\% of the most intense line - and rather featureless, the uncertainties of the original hyperfine constants can be safely used  to estimate the error bars of the recalculated ``observed'' frequencies, at least in the absence of crossovers in their fitting procedure. At this stage, the errors quoted in Table \ref{AJ} in the column ``this work'' are accuracy indicators that should be definitely refined through a final fit of the recorded spectra on the basis of the present analysis. Note that the relative intensity factors (\ref{rel_int}), useful to distinguish  the ``strong" from the ``weak" hyperfine components, are only used in the present work for building the spectra. They never affect however the equations allowing to extract the hyperfine parameters from the recalculated line frequencies.


\subsection{ $^{15}$N: $^{4}$P$_{5/2}\rightarrow ~^{4}$P$^{o}_{3/2,5/2}$ and $^{4}$P$_{5/2}\rightarrow ~^{4}$D$^{o}_{3/2,5/2,7/2}$}

The hyperfine spectra corresponding to the transitions $^{4}$P$_{5/2}\rightarrow ~^{4}$P$^{o}_{3/2,5/2}$ and $^{4}$P$_{5/2}\rightarrow ~^{4}$D$^{o}_{3/2,5/2,7/2}$ of $^{15}$N have two intense lines and one or two weak lines. The intense lines do not share any hyperfine level. Their frequency are given by the formula (\ref{nuhf}), where the quadrupole term vanishes ($I=1/2$). Let us set the frequency of the center of gravity of the  spectrum $\mathcal{S}$ to zero, and  define the $\mathcal{S}_c$ spectrum simulated with a new set of hyperfine constants that we want to determine. The $c$ subscript stands for the reassignment of the weak measured lines to crossovers. By defining  $\delta\nu_0$ as the center of gravity of this latter spectrum and denoting $\nu_{1}$ and $\nu_{2}$ the frequencies of two intense lines in the two $\mathcal{S}$ and $\mathcal{S}_c$ spectra, one has
\begin{eqnarray}
\nu_1=a'_1A'_e - a_1A_e &=& \delta\nu_0 + a'_1A' - a_1A \label{eq:15_1}\; ,\\
\nu_2=a'_2A'_e - a_2A_e &=& \delta\nu_0 + a'_2A' - a_2A \label{eq:15_2}
\end{eqnarray}
where $A'_e$ and $A_e$ ($A'$ and $A$) are respectively the hyperfine constants of the upper and lower levels, in the $\mathcal{S}$ ($\mathcal{S}_c$) spectrum.

The frequency $\nu_{3}$ of the experimental hyperfine line  interpreted as a crossover signal in the $\mathcal{S}_c$ spectrum and sharing the upper state with the intense line $\nu_{2}$, verifies
\begin{equation}
\nu_3=a'_2A'_e - a_1A_e = \delta\nu_0 + a'_2A' - \frac{a_1+a_2}{2}A\, . \label{eq:15_3}
\end{equation}
Equations (\ref{eq:15_1}), (\ref{eq:15_2}) and (\ref{eq:15_3}) form a well defined system of three linear equations for the unknowns $A'$, $A$ and $\delta\nu_0$. Solving it, we get
\begin{eqnarray}
A&=&2A_e\; , \label{x15N1}\\
A'&=&A'_e + \frac{a_1-a_2}{a'_1-a'_2}A_e\; , \label{x15N2}\\
\delta\nu_0&=&\frac{a'_1 a_2 - a_1a'_2}{a'_1-a'_2}A_e\; .\label{x15N3}
\end{eqnarray}
The new values of the  hyperfine constants ($A$ and $A'$) of the states involved in the transitions \mbox{$^{4}$P$_{5/2}\rightarrow ~^{4}$P$^{o}_{3/2,5/2}$} and \mbox{$^{4}$P$_{5/2}\rightarrow ~^{4}$D$^{o}_{3/2,5/2,7/2}$}, are presented in Table~\ref{AJ}. They are in good agreement with the  \emph{ab initio} values for all 
the considered states.

\begin{figure*}
\center
\resizebox{0.45\textwidth}{!}{\includegraphics{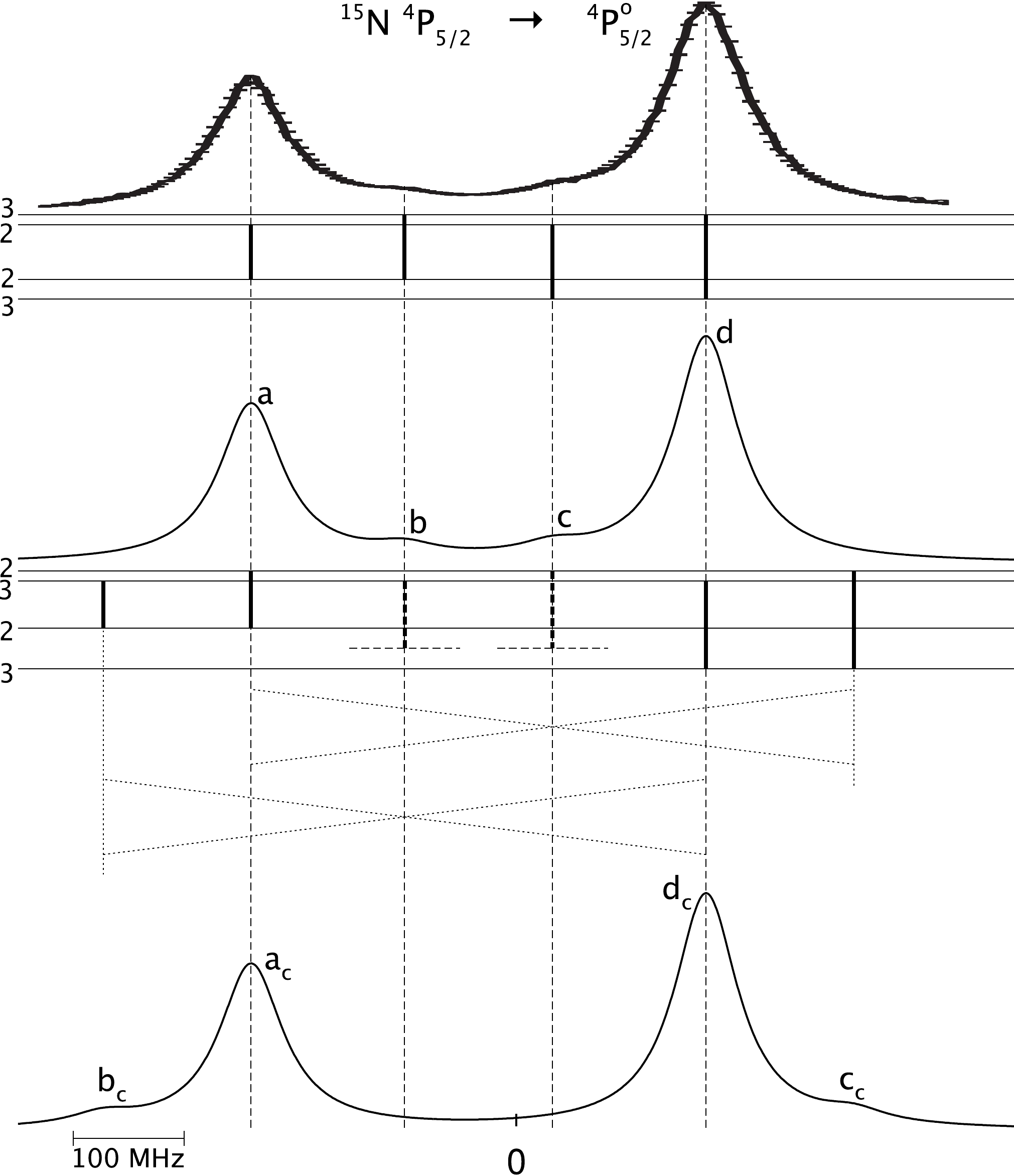}}
\resizebox{0.45\textwidth}{!}{\includegraphics{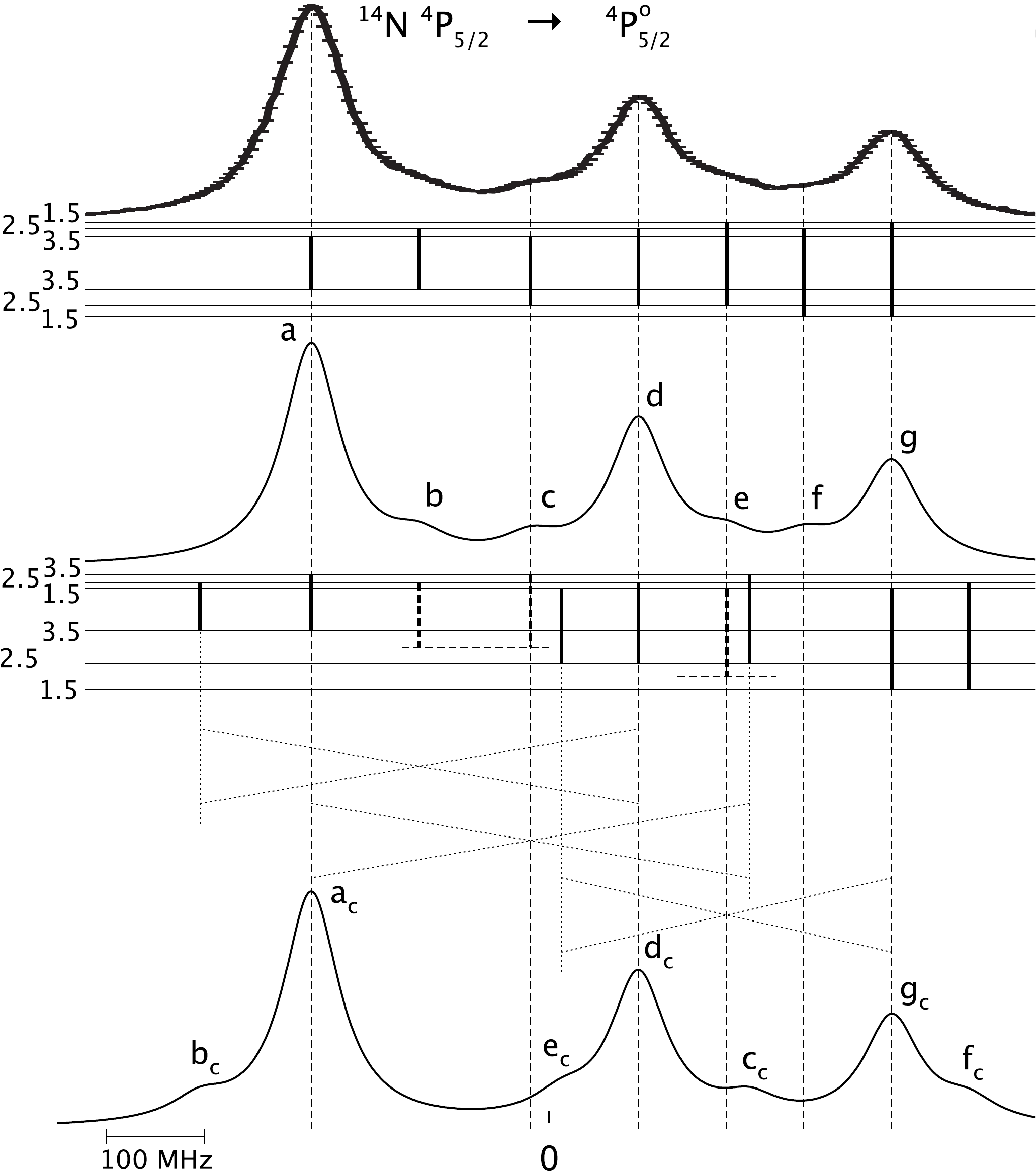}}
\caption{
Top : hyperfine spectra of the transition $^{4}$P$_{5/2}\rightarrow ~^{4}$P$^{o}_{5/2}$ recorded by Jennerich et al.~\cite{Jenetal:06a} for both isotopes (digitized from the figures of the original article). Middle:  level and transition diagrams and corresponding simulated  $\mathcal{S}$  spectra using the experimental hyperfine constants of Jennerich et al.~\cite{Jenetal:06a}. Bottom: level and transition diagrams and corresponding simulated  $\mathcal{S}_c$  spectra using the hyperfine constant values calculated from equations (9)-(11) for $^{15}$N and equations (17)-(21) for $^{14}$N. The crossovers, whose positions are indicated by the cross centers, are not included in the spectral synthesis.
The centers of gravity of  $\mathcal{S}_c$ and $\mathcal{S}$ are set to zero and to $ - \delta\nu_0$, respectively.
} \label{NDP1a}
\center
\resizebox{0.45\textwidth}{!}{\includegraphics{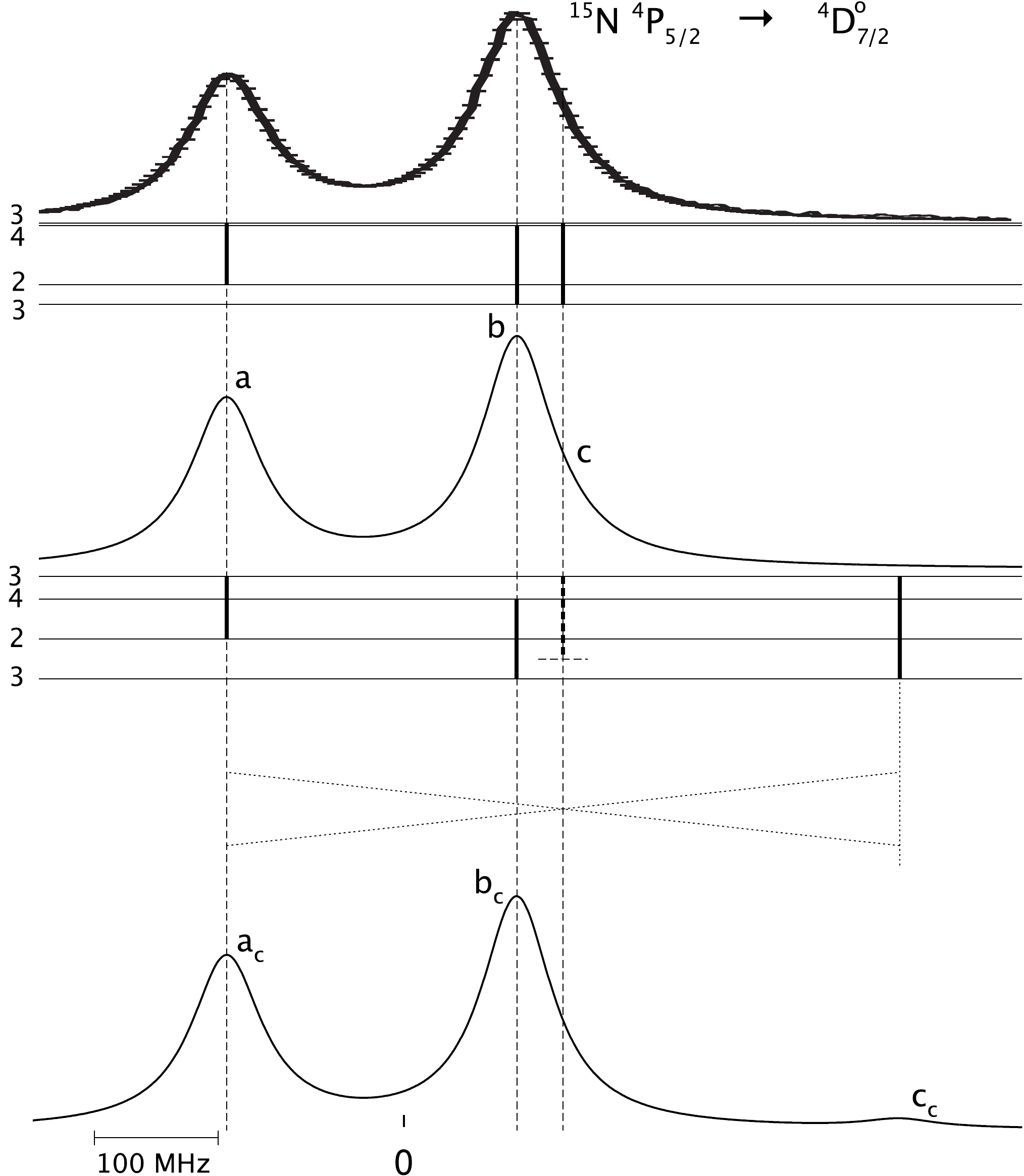}}
\resizebox{0.45\textwidth}{!}{\includegraphics{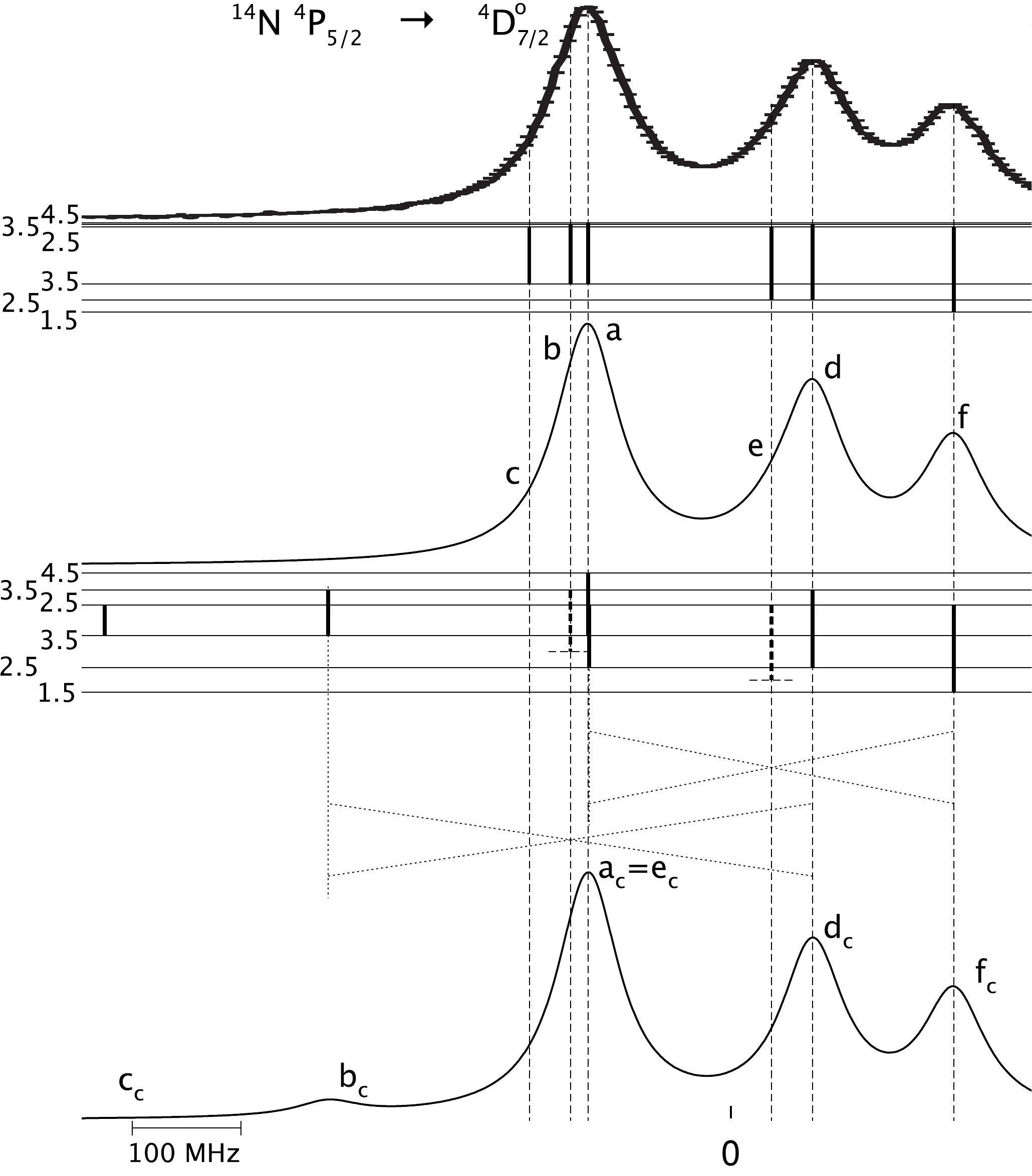}}
\caption{
Top : hyperfine spectra of the transition $^{4}$P$_{5/2}\rightarrow ~^{4}$D$^{o}_{7/2}$ recorded by Jennerich et al.~\cite{Jenetal:06a} for both isotopes (digitized from the figures of the original article). Middle:  level and transition diagrams and corresponding simulated  $\mathcal{S}$  spectra using the experimental hyperfine constants of Jennerich et al.~\cite{Jenetal:06a}. Bottom: level and transition diagrams and corresponding simulated  $\mathcal{S}_c$  spectra using the hyperfine constant values calculated from equations (9)-(11) for $^{15}$N and equations (17)-(21) for $^{14}$N. The crossovers, whose positions are indicated by the cross centers, are not included in the spectral synthesis. Note that $a_c=7/2 \rightarrow 9/2$ (100\%) and $e_c= 5/2 \rightarrow 5/2$ (6.5 \%) are superposed. The centers of gravity of  $\mathcal{S}_c$ and $\mathcal{S}$ are set to zero and to $ - \delta\nu_0$, respectively.
} \label{NDP1b}
\end{figure*}

The left parts of Figures~\ref{NDP1a} and \ref{NDP1b} display the recorded and simulated spectra for transitions $3s \; ^{4}$P$_{5/2}$ 
$ \rightarrow$ $~3p \; ^{4}$P$^{o}_{5/2},$ $^{4}$D$^{o}_{7/2}$ of $^{15}$N.
The uppermost spectrum is the one recorded by Jennerich~\emph{et al.}~\cite{Jenetal:06a}.
The middle ($\mathcal{S}$) and bottom ($\mathcal{S}_c$) synthetic spectra, with their corresponding level and transition diagrams,
 are calculated using the original experimental constants from \cite{Jenetal:06a} and the new constants $A'$ and $A$ derived from equations~(\ref{x15N1}) and (\ref{x15N2}), respectively.
We use the $c$-subscript to label the $\mathcal{S}_c$ lines. In the lower level diagrams, we represent our reassignment by indicating the concerned crossovers, including a thick dashed line linking the common upper level with a virtual level situated between the two involved lower levels. Furthermore, we draw a cross to emphasize the equidistance of the crossover from each of its underlying hyperfine transitions.
It should be pointed out that Jennerich et al.~\cite{Jenetal:06a} wrongly inverted the intensities of the hyperfine lines $b$ ($F=3 \rightarrow F'=4$) and $c$ ($F=3 \rightarrow F'=3$) in the level diagram of the transition $^{4}$P$_{5/2}\rightarrow ~^{4}$D$^{o}_{7/2}$. Moreover, line $c$ should have been labeled $b$ in Figure 2(g) of their article. Finally, the upper level of line $a$ should be $F'=3$ instead of $F'=4$ contrarily to what the hyperfine level diagram of their Figure~4(e) indicates.

From equation (\ref{x15N3}) we also deduce the center of gravity $\delta\nu_0$ of the different fine structure transitions. We obtain $\delta\nu_0=0$ for the transitions $^{4}$P$_{5/2}\rightarrow ~^{4}$P$^{o}_{5/2},~^{4}$D$^{o}_{5/2}$ ($J=J'$), $\delta\nu_0=-11.34(9)$MHz for $^{4}$P$_{5/2}\rightarrow ~^{4}$P$^{o}_{3/2},~^{4}$D$^{o}_{3/2}$ and $\delta\nu_0=5.67(5)$MHz for $^{4}$P$_{5/2}\rightarrow ~^{4}$D$^{o}_{7/2}$.

\subsection{ $^{14}$N: $^{4}$P$_{5/2}\rightarrow ~^{4}$P$^{o}_{3/2,5/2}$ and $^{4}$P$_{5/2}\rightarrow ~^{4}$D$^{o}_{3/2,5/2,7/2}$\label{sec:analyse14N}}

The case of isotope $^{14}$N is slightly complicated by the non-vanishing electric quadrupolar interaction, but the presence of three well identified lines permits us to perform the same analysis. If $\nu_1$, $\nu_2$ and $\nu_3$ are the intense line frequencies, their assignment gives~: 
\begin{eqnarray}
\nu_1&=&a'_1A'_e +b'_1B'_e - a_1A_e - b_1B_e\nonumber\\& =& \delta\nu_0 + a'_1A' + b'_1B' - a_1A - b_1B\; , \\
\nu_2&=&a'_2A'_e +b'_2B'_e - a_2A_e - b_2B_e\nonumber\\& =& \delta\nu_0 + a'_2A' + b'_2B' - a_2A - b_2B\; , \\
\nu_3&=&a'_3A'_e +b'_3B'_e - a_3A_e - b_3B_e\nonumber\\& =& \delta\nu_0 + a'_3A' + b'_3B' - a_3A - b_3B\; .
\end{eqnarray}
If amongst the observed weak lines of a given spectrum, $\nu_4$ and $\nu_5$ are identified as crossover signals, one has two additional constraints~:

\begin{eqnarray}
\nu_4&=&a'_2A'_e +b'_2B'_e - a_1A_e - b_1B_e\nonumber\\& =& \delta\nu_0 + a'_2A' + b'_2B' - \frac{a_1+a_2}{2}A - \frac{b_1+b_2}{2}B\; ,\\
\nu_5&=&a'_3A'_e +b'_3B'_e - a_2A_e - b_2B_e\nonumber\\& =& \delta\nu_0 + a'_3A' + b'_3B' - \frac{a_2+a_3}{2}A - \frac{b_2+b_3}{2}B \; .
\end{eqnarray}

As for isotope $^{15}$N, new experimental hyperfine constants $A'$, $A$, $B'$, $B$ and the center of gravity $\delta\nu_0$ of the considered transition are determined from
\begin{eqnarray}
A&=&2A_e\\
A'&=&A'_e - D^{-1}\nonumber\\ 
&&\hspace{-0.6cm}\times \left\{ A_e\left[ a_1(b_2'-b_3') + a_2(b_3'-b_1')+ a_3 (b_1' -  b_2') \right]\right.\nonumber\\
&&\hspace{-0.6cm}\left. + B_e \left[ b_1(b_2'-b_3') + b_2(b_3'-b_1')+ b_3 (b_1' -  b_2') \right]\right\}\\
B&=&2B_e\\
B'&=&B'_e - D^{-1}\nonumber\\
&&\hspace{-0.6cm}\times \left\{ A_e\left[ a_1(a_2'-a_3') + a_2(a_3'-a_1')+ a_3 (a_1' - a_2') \right]\right.\nonumber\\
&&\hspace{-0.6cm}\left. +  B_e \left[ b_1(a_2'-a_3') + b_2(a_3'-a_1')+ b_3 (a_1' -  a_2') \right]\right\}\\
\delta\nu_0&=&\frac{\alpha A_e + \beta B_e}{D}
\end{eqnarray}
where
\begin{eqnarray*}
D &=& a'_1(b_2'-b_3') + a'_2(b_3'-b_1')+ a'_3 (b_1' -  b_2')\; ,\\
\alpha &=& a_1(a_2'b_3'-a_3'b_2') + a_2(a_3'b_1'-a_1'b_3')+ a_3 (a_1'b_2' - a_2' b_1')\; ,\\
\beta &=& b_1(a_2'b_3'-a_3'b_2') + b_2(a_3'b_1'-a_1'b_3')+ b_3 (a_1'b_2' - a_2' b_1')  \, .
\end{eqnarray*}

\begin{figure*}
\center
\resizebox{0.46\textwidth}{!}{\includegraphics{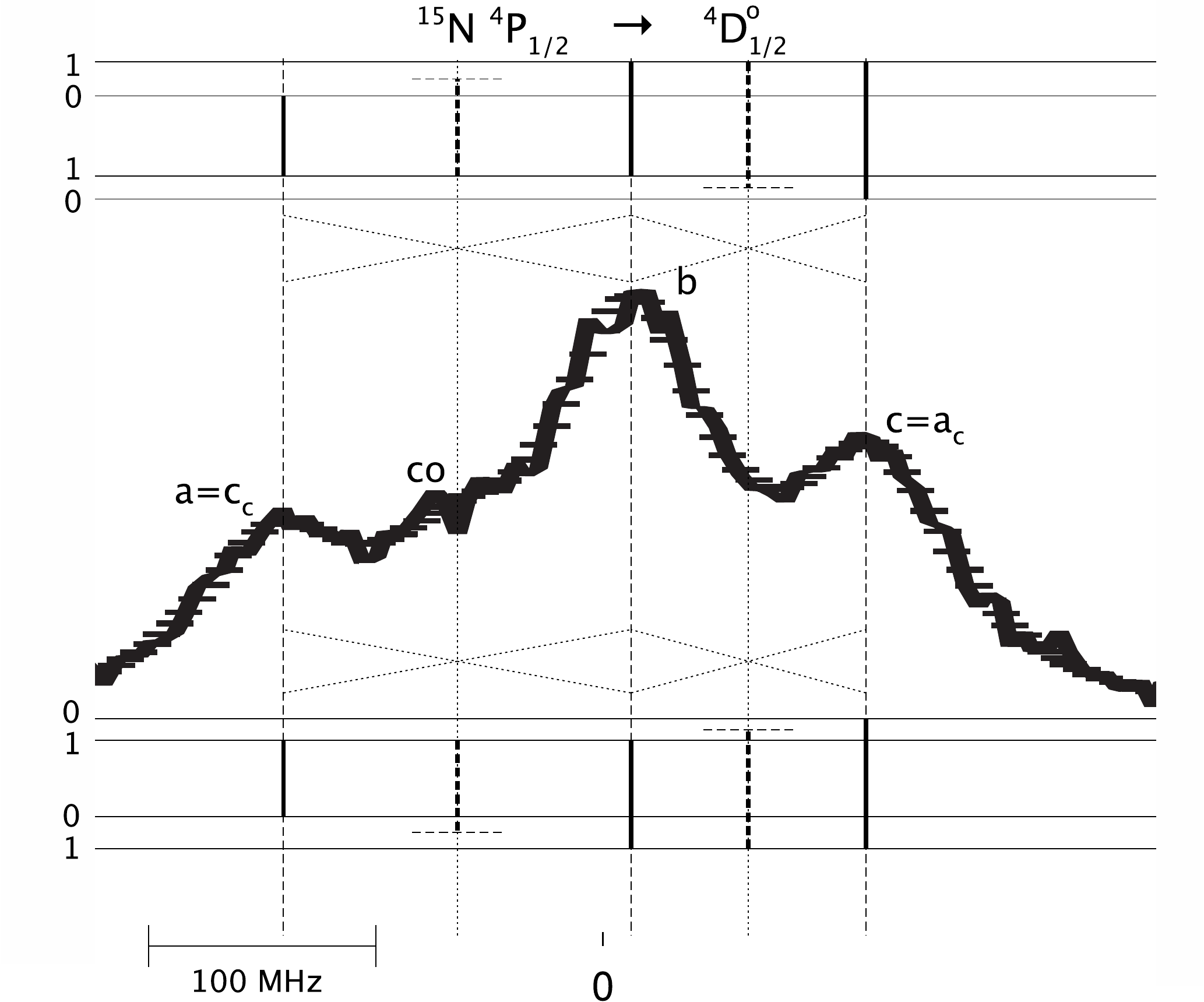}}
\resizebox{0.46\textwidth}{!}{\includegraphics{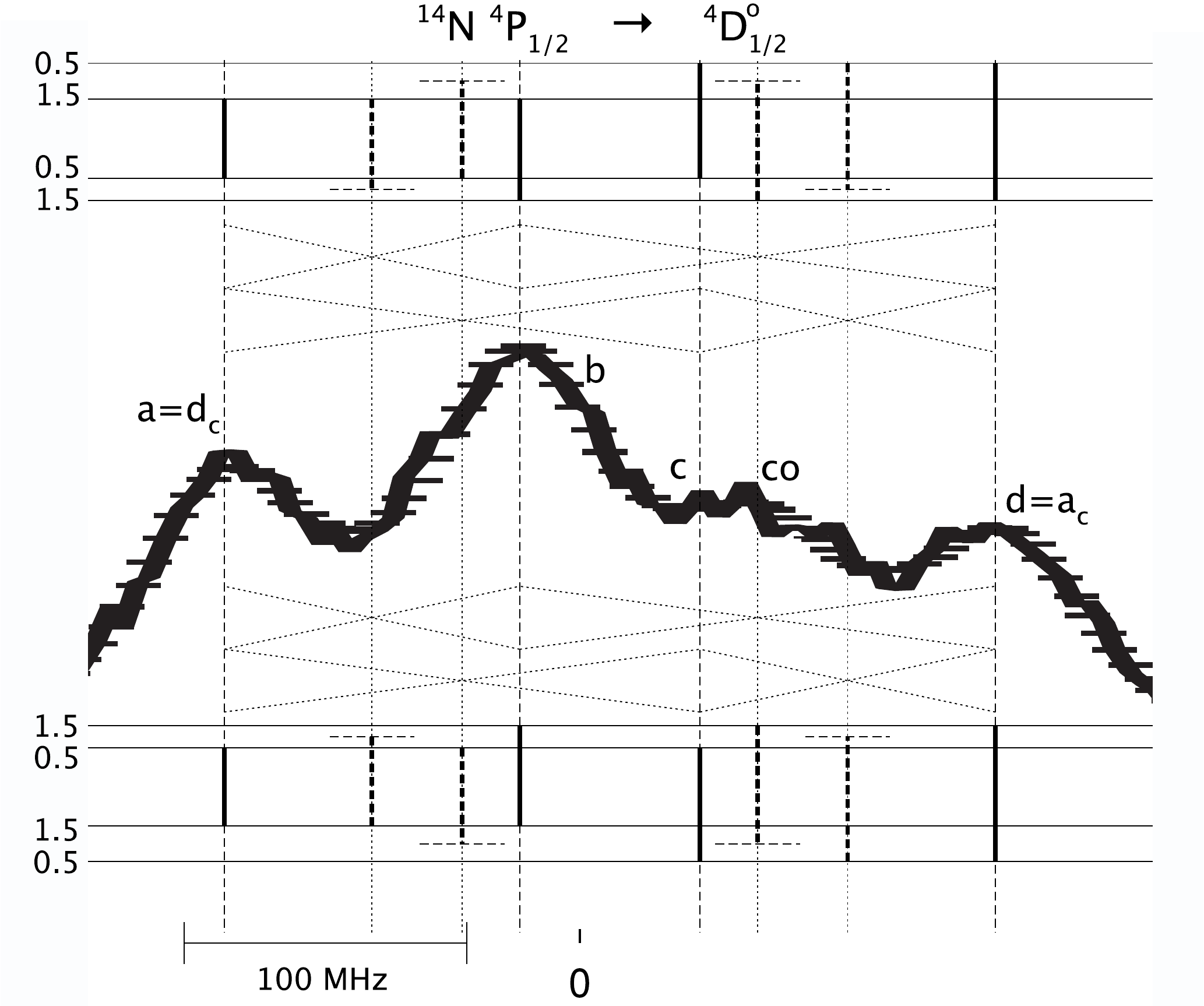}}
\caption{
Middle : hyperfine spectra of the transition $^{4}$P$_{1/2}\rightarrow~^{4}$D$^{o}_{1/2}$ recorded by Jennerich et al.~\cite{Jenetal:06a} for both isotopes (digitized from the figures of the original article). Top and bottom :  level and transition diagrams representing the two assignments proposed by Jennerich et al.~\cite{Jenetal:06a}. The top assignments are the ones favoured by Jennerich et al. We definitely opt for the bottom assignments, with crossover  (co)  signals involving lines sharing their upper level, on the basis of a much better agreement with theory \cite{Jonetal:10a}. The center of gravity of the spectrum is set to zero.
} \label{NDP2}
\end{figure*}

The so-deduced values of $A'$, $A$, $B'$ and $B$ are given in Table~\ref{AJ}. They are in good agreement with the {\em ab initio} theoretical hyperfine constants. Like above, the right parts of Figures \ref{NDP1a} and \ref{NDP1b} present the hyperfine spectra and level diagrams of the two transitions $^{4}$P$_{5/2}\rightarrow ~^{4}$P$^{o}_{5/2}$, $^{4}$D$^{o}_{7/2}$ for the isotope $^{14}$N. 
We should again point out that the Figure~3(e) of Jennerich et al.~\cite{Jenetal:06a} is misleading since it shows an increasing energy for a decreasing $F$ for the $^4D^o_{7/2}$. It corresponds to a negative $A_{7/2}(^4D^o)$ while it is, according to their analysis, positive. Their systematic labeling of the lines ($a, b, c, \dots$ for increasing hyperfine transition energy) is therefore not respected for this spectrum.

We find $\delta\nu_0=0$ for the transitions \mbox{$^{4}$P$_{5/2}\rightarrow ~^{4}$P$^{o}_{5/2},$} $^{4}$D$^{o}_{5/2}, \delta\nu_0=21.68(17)$~MHz for \mbox{$^{4}$P$_{5/2}\rightarrow ~^{4}$P$^{o}_{3/2},$} $^{4}$D$^{o}_{3/2}$ and \mbox{$\delta\nu_0=-10.84(9)$MHz} for \mbox{$^{4}$P$_{5/2}\rightarrow ~^{4}$D$^{o}_{7/2}$}.

One should expect the quadrupole hyperfine constant $B$ of the state $^{4}$D$^{o}_{3/2}$ to be rather small. Indeed, in the context of hyperfine simulations based on the Casimir formula and in the non-relativistic approximation, the main contribution to the $B$ constants is given by~\cite{Hib:75b}~:
\begin{equation}\label{eq:B}
B_{J}=-G\,Q\,b_{q}\,\frac{6\langle\,\vec{L}.\vec{J}\,\rangle^{2}\,-\,3\langle\,\vec{L}.\vec{J}\,\rangle\,-\,2L(L+1)J(J+1)}{L(2L-1)(J+1)(2J+3)}
\end{equation}
where $G=234.96475$ is used for obtaining $B_J$ in MHz when expressing the quadrupole moment $Q$  in barns, the hyperfine parameter $b_{q}$ in $a_0^{-3}$ and with
\begin{equation}
\langle\,\vec{L}.\vec{J}\,\rangle = \frac{1}{2}\left[ J(J+1) + L(L+1) - S(S+1) \right] \,.
\end{equation}
 It is easily verified that equation (\ref{eq:B}) vanishes for the $^{4}$D$^{o}_{3/2}$ state of $^{14}$N. Therefore, a non-zero $B($$^{4}$D$^{o}_{3/2})$ value should be interpreted as arising from higher order relativistic effects and/or in terms of hyperfine interaction between levels of different $J$. From the present analysis, we deduce $B($$^{4}$D$^{o}_{3/2})=-0.9(17)$~MHz, which is from this point of view, more realistic than the experimental value ($+10.46(88)$~MHz) reported in~\cite{Jenetal:06a}.

\subsection{The transition  $^{4}$P$_{1/2}\rightarrow ~^{4}$D$^{o}_{1/2}$}

The spectrum of the transition \mbox{$^{4}$P$_{1/2}\rightarrow ~^{4}$D$^{o}_{1/2}$} of  isotope $^{15}$N is resolved (Figure~\ref{NDP2}) but the lines $a$ ($F=1 \rightarrow F'=0$) and $c$ ($F=0 \rightarrow F'=1$) have the same relative intensities (50\% of the most intense line). It causes an \emph{a priori} ambiguous assignment. The same problem appears in the  $^{14}$N spectrum where lines $a$ ($F=1/2 \rightarrow F'=3/2$) and $d$ ($F=3/2 \rightarrow F'=1/2$) have the same relative intensities (80\% of the most intense line). Line $c$ ($F=1/2 \rightarrow F'=1/2$, 10\%) is not helpful since it does not appear in this spectrum. Because of this identification problem, Jennerich et al.~\cite{Jenetal:06a} suggested two possible values for each of the hyperfine constants of the $^{4}$P$_{1/2}$ and $^{4}$D$^{o}_{1/2}$ states (see Table~\ref{AJ}). The first proposition corresponds to the case where  $\nu_a < \nu_c$ for $^{15}$N and  $\nu_a < \nu_d $ for $^{14}$N (upper level diagram of Fig. \ref{NDP2}). The second proposition corresponds to the inverse situation (lower level diagram of Fig. \ref{NDP2}). On the basis of the presence of a crossover with a  positive intensity between lines $b$ and $d$ in the spectrum of $^{14}$N, Jennerich et al.~\cite{Jenetal:06a} estimated that the first proposition is the most likely. Indeed, they infer from the presence of this crossover that $b$ and $d$ share their lower level, leading to the identification of the lines $a$ and $d$. The same argument was used for $^{15}$N using the crossover between $a$ and $b$. A similar argument has been used in previous studies of Chlorine~\cite{TatWal:99a} and Oxygen~\cite{JenTat:00a}. However, a crossover arising from two transitions with a common {\it upper} level may also have a positive intensity~\cite{Hanetal:71a,Andetal:78a}. Combining this observation with the fact that the agreement between the hyperfine constants of the states $^{4}$P$_{1/2}$ and $^{4}$D$^{o}_{1/2}$ and the {\em ab initio} values~\cite{Jonetal:10a} is much better with the second set than with the first one, we think that the first choice of Jennerich et al.~\cite{Jenetal:06a} is not the good one, and we definitely opt for the second one.

\begin{figure*}
\center
\resizebox{0.45\textwidth}{!}{\includegraphics{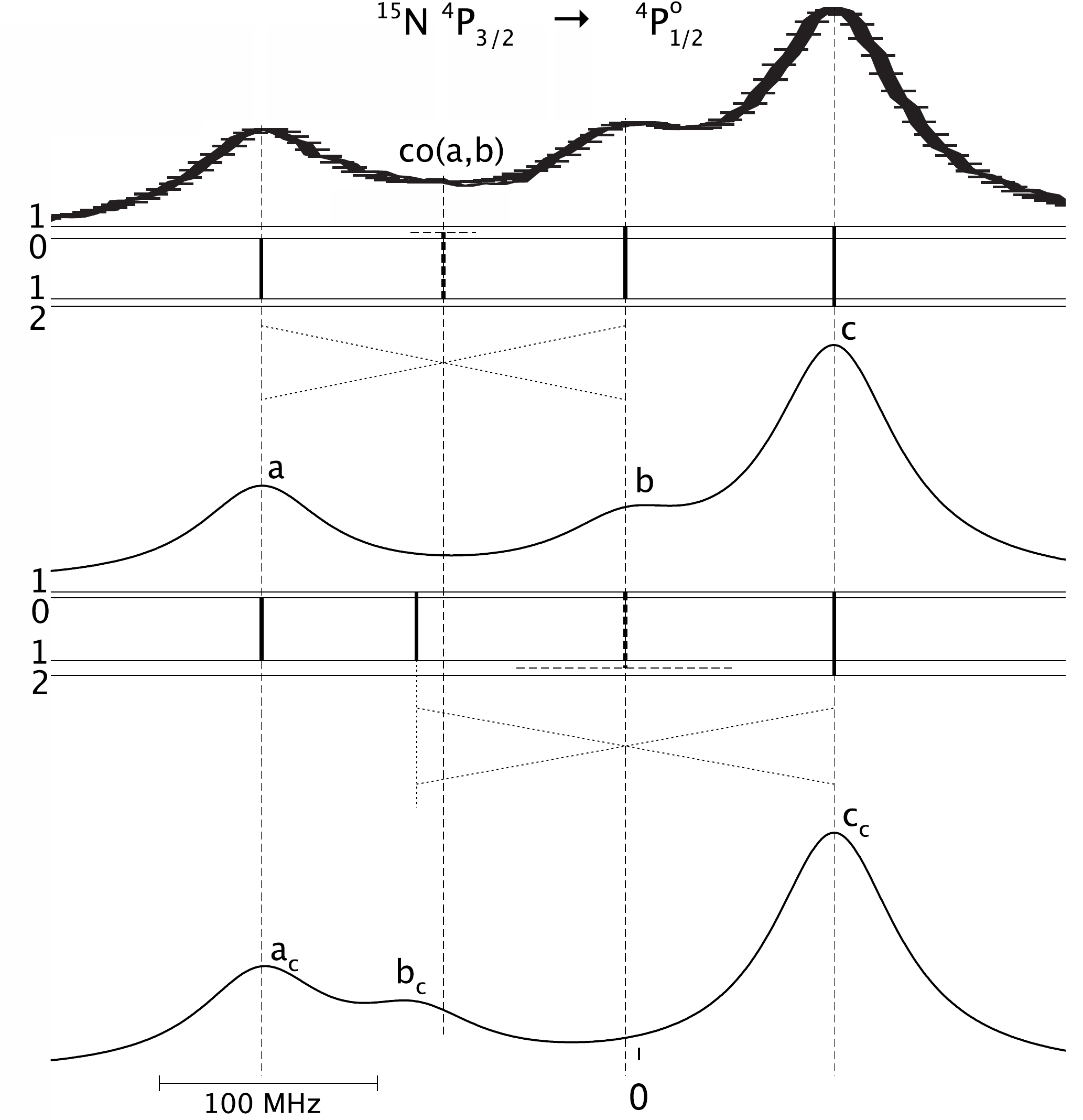}}
\resizebox{0.45\textwidth}{!}{\includegraphics{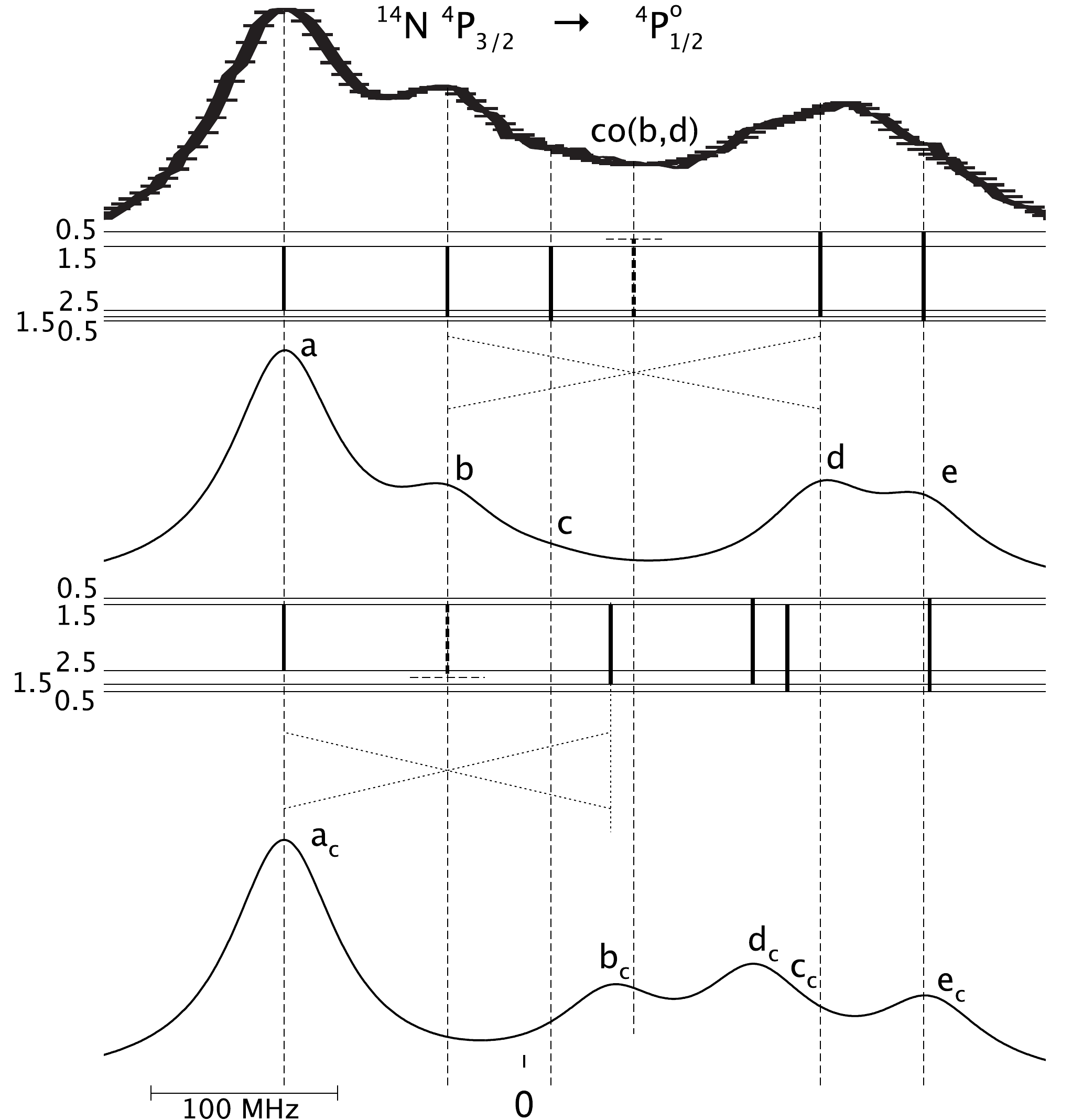}}
\caption{
Top : hyperfine spectra of the transition $^{4}$P$_{3/2}\rightarrow ~^{4}$P$^{o}_{1/2}$ recorded by Jennerich et al.~\cite{Jenetal:06a} for both isotopes (digitized from the figures of the original article). Middle :  level and transition diagrams and corresponding simulated  $\mathcal{S}$  spectra using the experimental hyperfine constants of Jennerich et al.~\cite{Jenetal:06a}. Bottom : level and transition diagrams and corresponding simulated  $\mathcal{S}_c$  spectra using the hyperfine constant values calculated from equations (9)-(11) for $^{15}$N and equations (28)-(31) for $^{14}$N. We reassigne the crossovers identified by Jennerich et al. and marked in the experimental spectra as co($a,b$) and co($b,d$) for $^{15}$N and $^{14}$N respectively, as real hyperfine components ($b_c$ in the $\mathcal{S}_c$  spectra). The crossovers are not included in the spectral synthesis. The centers of gravity of  $\mathcal{S}_c$ and $\mathcal{S}$ are set to zero and to $ - \delta\nu_0$, respectively.
} \label{NDP3}
\end{figure*}

\subsection{The transition $^{4}$P$_{3/2}\rightarrow ~^{4}$P$^{o}_{1/2}$}

The recorded transition spectra of $\; ^{4}$P$_{3/2}\rightarrow \; ^{4}$P$^{o}_{1/2}$ is showed at the top left of Figure \ref{NDP3}. For the $^{15}$N spectrum, with two well identified lines ($a$ and $c$) and an experimental line $b$ that we interpret as a crossover signal of $b_c$ and $c_c$, we applied the same procedure described above, using equations~(\ref{x15N1})-(\ref{x15N3}). The new constants $A'=A(~^{4}$P$^{o}_{1/2})$ and $A=A(~^{4}$P$_{3/2})$ that we infer are reported in Table~\ref{AJ} ($\delta\nu_0=-11.98(12)$~MHz). Simulated spectra and corresponding level and transition diagrams are presented in the left part of Figure~\ref{NDP3}. The experimental set of hyperfine constants of Jennerich et al.~\cite{Jenetal:06a} and the present one generate simulated spectra, $\mathcal{S}$ and $\mathcal{S}_c$, that do not agree as well as for the above discussed transitions. 
If we  reinterpret the crossover co($a,b$) suggested by Jennerich et al. as a real line ($b_c$, $F=1 \rightarrow F'=1$)\footnote{The corresponding line intensity factor calculated from equation~(\ref{rel_int}), 20\% of the most intense line, would suggest a slightly stronger signal than observed.}, the relative disagreement could be attributed to the fact that  the line $b_c$   is strong enough to perturb deeply the $a$ line shape. For this effect, we refer to the section 4.2 of Jennerich et al.~\cite{Jenetal:06a} who discussed the possibility of observing line shape perturbation in some transitions, in particular when the separation of two lines is comparable to the natural linewidth. Furthermore, the possible presence of a crossover between $a_c$ and $b_c$, and the deviation of co($a,b$) from $b_c$ could affect the quality of Jennerich et al.'s fit, making too optimistic the uncertainty of their  experimental hyperfine constants that we use for building the $\mathcal{S}_c$ spectrum.

This problem appears even more seriously in the case of $^{14}$N. Indeed, the transition  $^4$P$_{3/2} \rightarrow ^4$P$^o_{1/2}$ cannot be analysed according to section \ref{sec:analyse14N}. Its hyperfine spectrum is composed of only one well identified line ($a$:100\%, $ \nu_1$ $=5/2$ $\rightarrow$ $3/2$), three nearly equally intense lines ($b$:30\%, $\nu_2$=3/2$\rightarrow$3/2; $d$:37\%, $\nu_4$=3/2$\rightarrow$1/2; $e$:30\%, $\nu_5$=1/2$\rightarrow$1/2), one line which is too weak to be visible ($c$:3.7\%, $\nu_3$=1/2 $ \rightarrow$ 3/2) and possibly many crossovers. Trusting our line assignment for the same spectrum in $^{15}$N, it is unlikely that the experimental line $b$ could be anything else than a crossover of the most intense line ($a_c$) and another hyperfine transition. The best candidate for the latter is $b_c$. We suppose
\begin{eqnarray}
\nu_1&=&a'_1A'_e +b'_1B'_e - a_1A_e - b_1B_e\nonumber\\& =& \delta\nu_0 + a'_1A' + b'_1B' - a_1A - b_1B\; , \\
\nu_2&=&a'_1A'_e +b'_1B'_e - a_2A_e - b_2B_e\nonumber\\& =& \delta\nu_0 + a'_1A' + b'_1B' - \frac{a_1+a_2}{2}A - \frac{b_1+b_2}{2}B.
\end{eqnarray}
On the other hand, the transition $\nu_3$=1/2$\rightarrow$3/2 is too weak to be observed and the identification of the transitions $\nu_4$=3/2$\rightarrow$1/2 et $\nu_5$=1/2$\rightarrow$1/2 are uncertain. The observed peak in the region of those lines is interpreted as their superposition with their crossovers. It is therefore impossible to extract the hyperfine constants $A$ and $A'$ from the $^{14}$N  experimental data alone.

Nonetheless, neglecting in the simulation a crossover signal between these transitions introduces an error on the $\nu_4$ and $\nu_5$ lines that we denote respectively $\epsilon_{\nu_4}$ and $\epsilon_{\nu_5}$. We then have
\begin{eqnarray}
\nu_4&=&\epsilon_{\nu_4} + a'_2A'_e - a_2A_e - b_2B_e \nonumber\\& =& \delta\nu_0 + a'_2A'  - a_2A - b_2B \\
\nu_5&=&\epsilon_{\nu_5}+a'_2A'_e - a_3A_e - b_3B_e \nonumber\\& =& \delta\nu_0 + a'_2A'  - a_3A - b_3B
\end{eqnarray}
These four constraints permit to express the hyperfine constants involved in this transition as a  function of $\epsilon_{\nu_4}$ and $\epsilon_{\nu_5}$~:
\begin{eqnarray}
A'= A_{1/2}(^4 \textrm{P}^o)&=&-74.8(34) - \frac{2}{3} \epsilon_{\nu_4} \\
A= A_{3/2}(^4 \textrm{P}) &=&\phantom{-} 61.79(95) - \frac{1}{6} ( \epsilon_{\nu_4} -\epsilon_{\nu_5} )\\
B= B_{3/2}(^4 \textrm{P}) &=&\phantom{-} 16.54(82) + \frac{1}{3} (\epsilon_{\nu_4} - \epsilon_{\nu_5})\\
\delta\nu_0&=&\phantom{-} 14.60(29) + \frac{1}{6} (\epsilon_{\nu_4} + \epsilon_{\nu_5} )
\end{eqnarray}
If we impose A$(^{15}$N$)=-1.4028$ A$(^{14}$N) to get $A$ and $A'$, we find $B_{3/2}(^4 \textrm{P})=3.5(91)$ MHz, $\epsilon_{\nu_4}=-36.0(75)$~MHz, $\epsilon_{\nu_5}=3(18)$ MHz and $\delta\nu_0=9.1(44)$~MHz. Figure~\ref{NDP3} displays the corresponding $\mathcal{S}_c$ simulated spectrum.

\begin{figure}
\center 
\resizebox{0.49\textwidth}{!}{\includegraphics{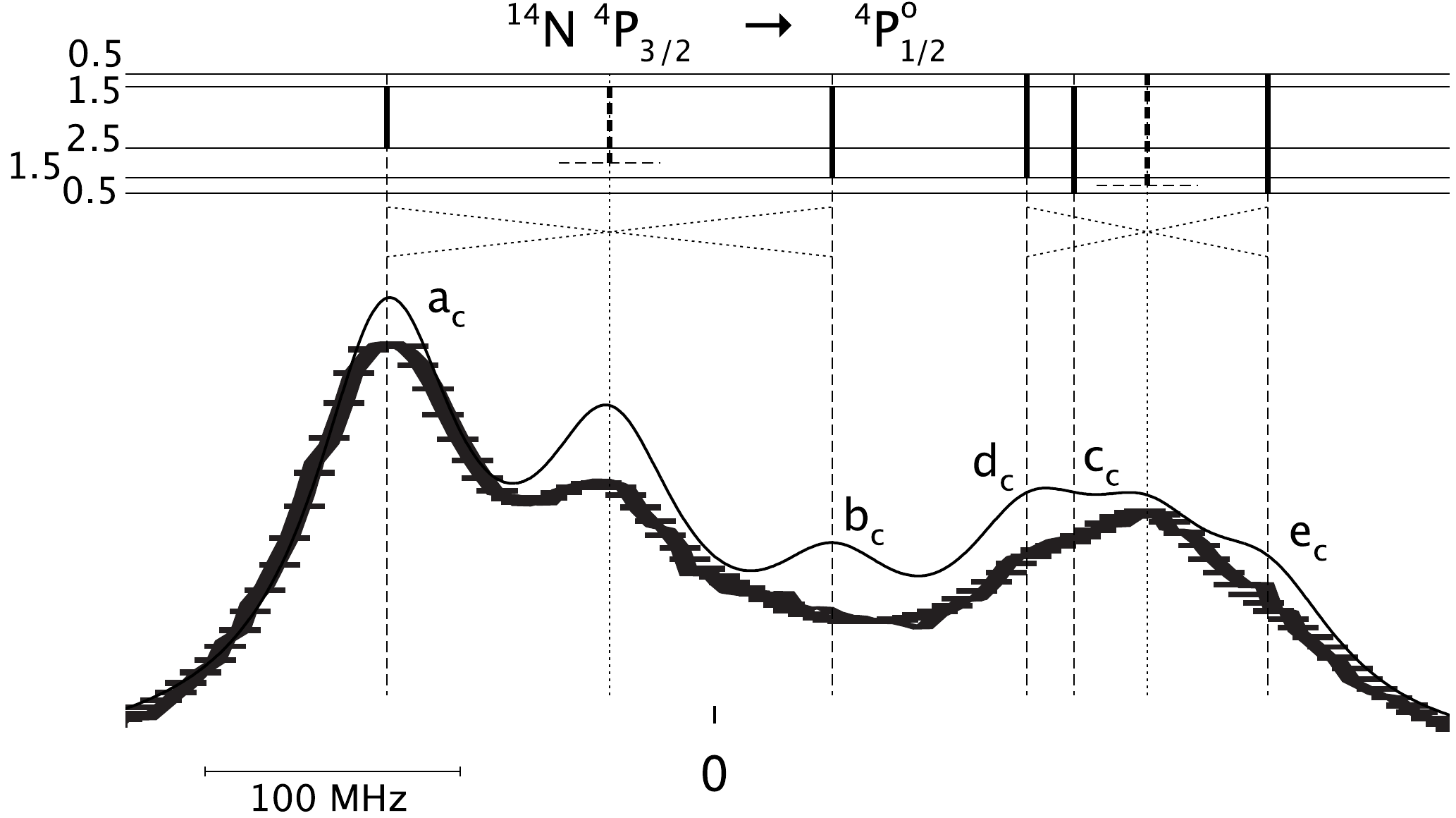}}
\caption{Comparison of the recorded spectrum of \cite{Jenetal:06a} with the $\mathcal{S}_c$ simulation for the $^{4}$P$_{3/2}\rightarrow ~^{4}$P$^{o}_{1/2}$ transition of the $^{14}$N. Crossovers co($a,b$) and co($d,e$), indicated in the level and transition diagram, are included in the simulation, with an intensity equal to the mean of the two interfering transitions. 
For the sake of clarity, the vertical position of the simulated and recorded spectra are shifted re shifted with respect to each other.
 The center of gravity of the spectrum is set to zero.
} \label{co}
\end{figure}

In Figure~\ref{co}, we finally tempt a crude spectral synthesis including the two crossover signals co$(a,b)$ and co$(d,e)$. The intensity values are estimated from the average of the two hyperfine relative intensities calculated according to equation~(\ref{rel_int}). The  $\mathcal{S}_c$ spectrum compares relatively well with the recorded one but for the disappearance, as in the previously analyzed spectra, of the reassigned transition~($b_c$).

\subsection{Transitions $^{4}$P$_{3/2}\rightarrow ~^{4}$P$^{o}_{5/2}$ and $^{4}$P$_{1/2}\rightarrow ~^{4}$P$^{o}_{3/2}$}\label{sec:nRes}

\begin{figure*}
\center
\resizebox{0.9\textwidth}{!}{\includegraphics{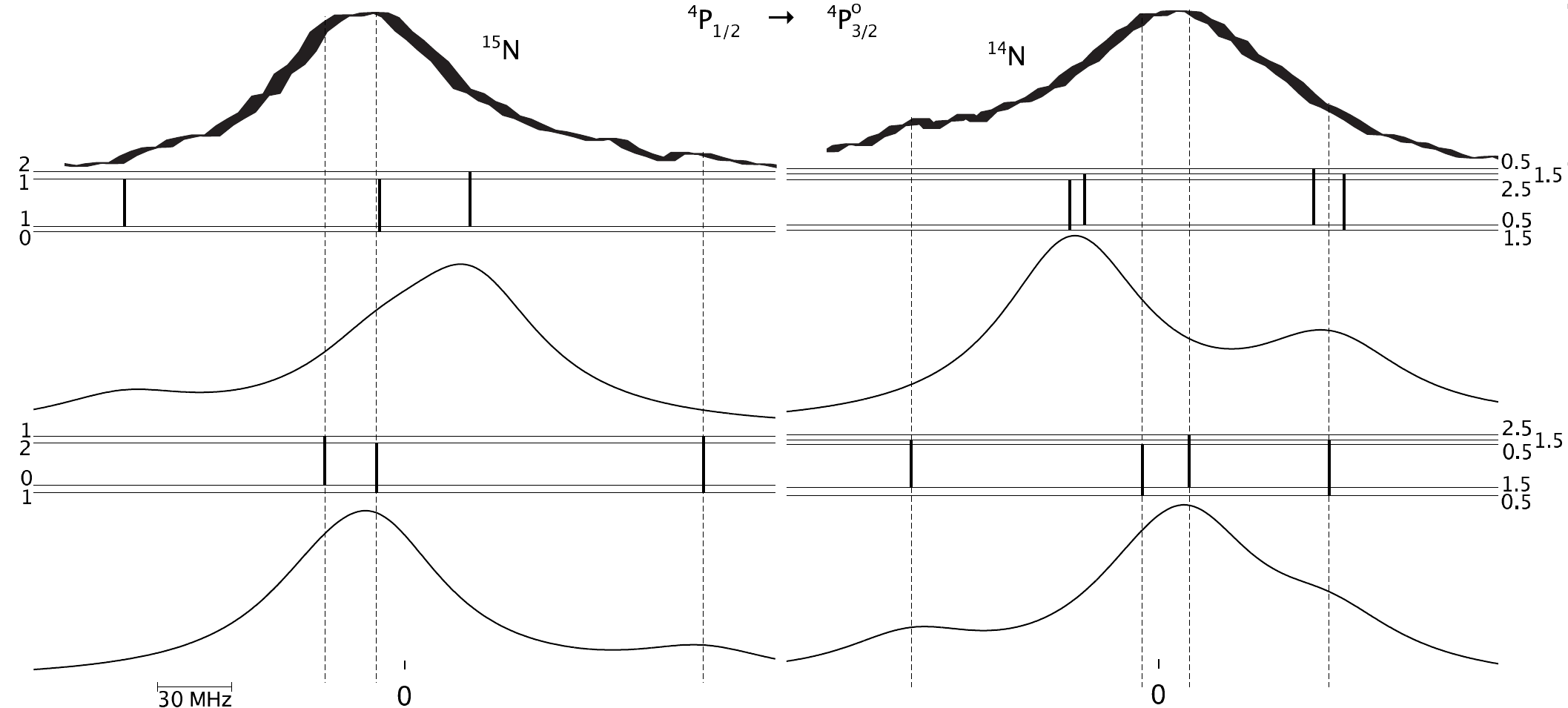}}
\caption{
Top : hyperfine spectra of the transition $^{4}$P$_{1/2}\rightarrow^{4}$P$^{o}_{3/2}$ recorded by Jennerich et al.~\cite{Jenetal:06a} for both isotopes. Middle :  the $\mathcal{S}$ simulated spectra using the experimental hyperfine constants of Jennerich et al.~\cite{Jenetal:06a}. Bottom : the $\mathcal{S}_c$ simulated spectra with the corresponding level and transition diagrams, calculated with the present hyperfine structure constants. The scale of the recorded spectra is adjusted with respect to the $\mathcal{S}_c$ simulation simultaneously for both isotopes, using the most intense peak of $^{14}$N and the small signal to its left that we assign to the $F=3/2 \rightarrow F'=3/2$ hyperfine component. The center of gravity of $\mathcal{S}$ and $\mathcal{S}_c$ is set to zero.
} \label{nRes1a}
\end{figure*}

\begin{figure*}
\center
\resizebox{0.9\textwidth}{!}{\includegraphics{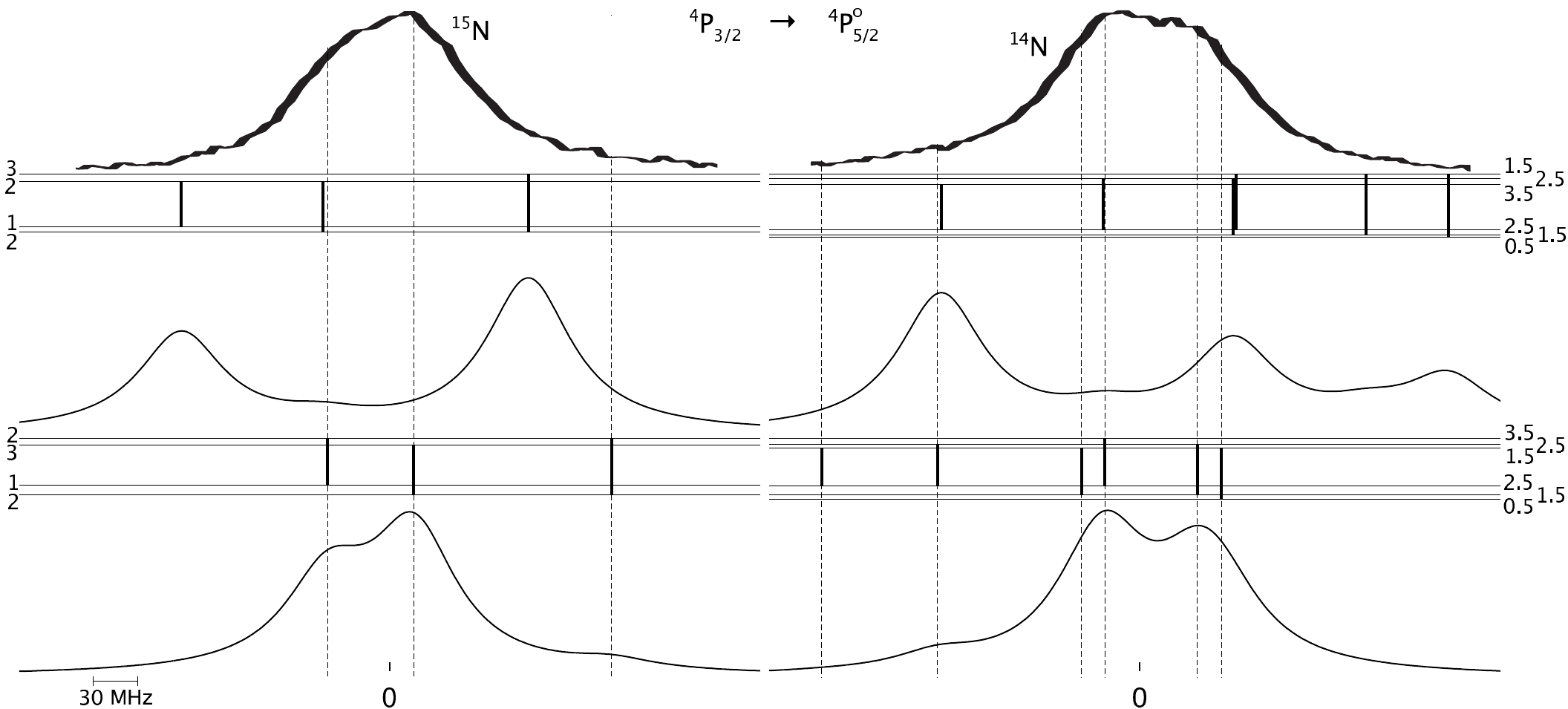}}
\caption{
Top : hyperfine spectra of the transition$^{4}$P$_{3/2}\rightarrow^{4}$P$^{o}_{5/2}$  recorded by Jennerich et al.~\cite{Jenetal:06a} for both isotopes. Middle :  the $\mathcal{S}$ simulated spectra using the experimental hyperfine constants of Jennerich et al.~\cite{Jenetal:06a}. Bottom : the $\mathcal{S}_c$ simulated spectra with the corresponding level and transition diagrams, calculated with the present hyperfine structure constants.  The scale of the recorded spectra is adjusted with respect to the $\mathcal{S}_c$ simulation simultaneously for both isotopes. The center of gravity of $\mathcal{S}$ and $\mathcal{S}_c$ is set to zero.
} \label{nRes1b}
\end{figure*}

The hyperfine spectra of the transitions \mbox{$^{4}$P$_{3/2}\rightarrow ~^{4}$P$^{o}_{5/2}$} and \mbox{$^{4}$P$_{1/2}\rightarrow ~^{4}$P$^{o}_{3/2}$} were also recorded by Jennerich et al.~\cite{Jenetal:06a}. Our simulated spectra are compared with the measured ones in Figures \ref{nRes1a} and \ref{nRes1b}. The $\mathcal{S}$ spectrum of the transition  $^{4}$P$_{3/2}\rightarrow ~^{4}$P$^{o}_{5/2}$ simulated with the experimental hyperfine constants determined by Jennerich et al.~\cite{Jenetal:06a} indicates that the measured spectrum should be resolved while it is not in reality. There is another contradiction between the synthetic and experimental spectra for the transition $^{4}$P$_{1/2}\rightarrow ~^{4}$P$^{o}_{3/2}$ of isotope $^{14}$N. There is indeed an asymmetry in the measured line that suggests the presence of a low intensity line to the left of it (see the uppermost spectrum of the right part of Figure~\ref{nRes1a}), while the experimental hyperfine constant set tends to predict it to the right. To explain this discrepancy between simulations and observations, the authors suggested that strong line shape perturbation could appear in these transitions in a saturation spectroscopy experiment (cf. section 4.2 of \cite{Jenetal:06a}).
To the contrary, the simulations based on the present reinterpretation (bottom of Figures~\ref{nRes1a}) are in agreement with the non-resolved spectra and small features in those lines are assigned.

\begin{figure*}
\center
\resizebox{0.45\textwidth}{!}{\includegraphics{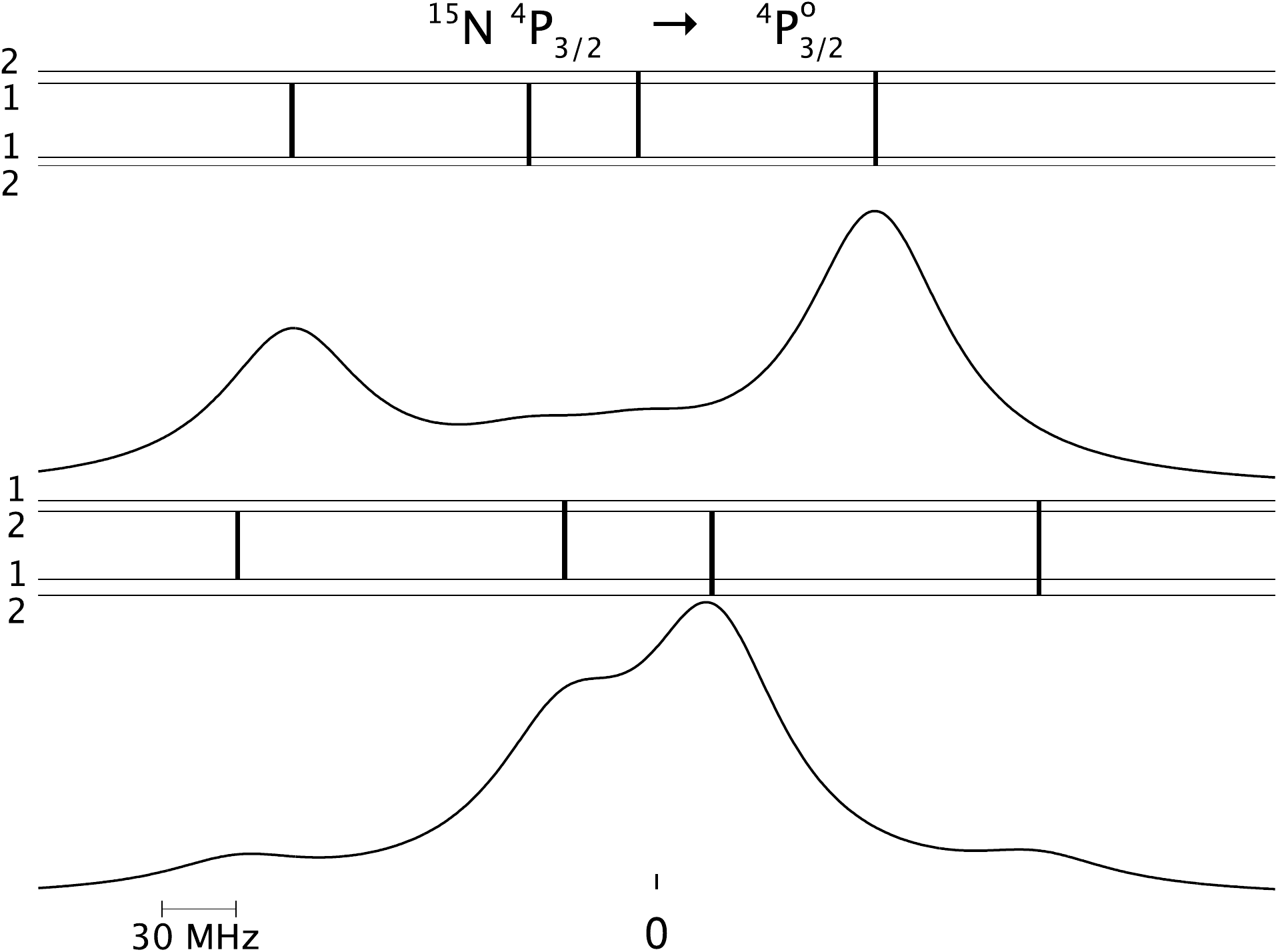}}
\resizebox{0.45\textwidth}{!}{\includegraphics{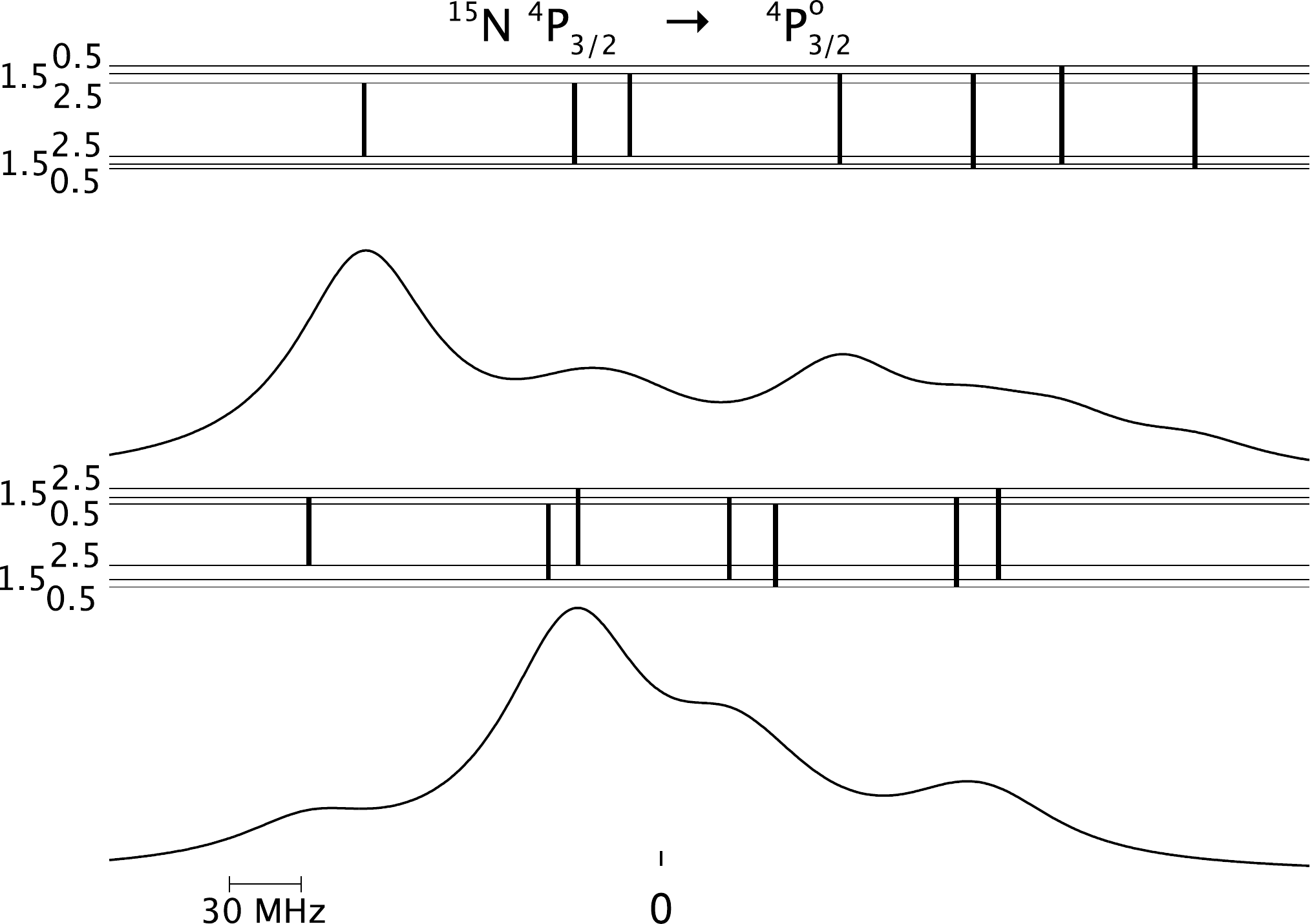}}
\caption{
Hyperfine spectra simulations of the transition $^{4}$P$_{3/2}\rightarrow^{4}$P$^{o}_{3/2}$ for both isotopes. Top : line and transition diagram with the corresponding $\mathcal{S}$ simulated spectra using the experimental hyperfine constants of Jennerich et al.~\cite{Jenetal:06a}. Bottom :  line and transition diagram with the corresponding $\mathcal{S}_c$ simulated spectra using the present hyperfine structure constants.  The center of gravity of $\mathcal{S}$ and $\mathcal{S}_c$ is set to zero.
} \label{nRes2}
\end{figure*}

The only measurement of the spectra of the transition $^{4}$P$_{3/2}\rightarrow ~^{4}$P$^{o}_{3/2}$ was recorded by Cangiano et al. \cite{Canetal:94a} but no figure is presented for it.
Figure \ref{nRes2} displays both $\mathcal{S}$ and $\mathcal{S}_c$ simulated spectra and corresponding level and transition diagrams using respectively Jennerich et al.'s experimental (top) and present (bottom) sets of hyperfine constants. The resulting spectra are respectively resolved and unresolved.


\section{$J$-dependent specific mass shifts in \mbox{$3s~^{4}P\rightarrow 3p~^{4}L^{o}$} transitions}

The isotope shift (IS) of a transition is often separated in three contributions : the normal mass shift (NMS), linear in  the line frequency, the specific mass shift (SMS), which is proportional to the change of the mass polarization term expectation value  between the two levels involved in the transition
\begin{equation}
\Delta \left \langle \sum_{i<j} {\bf p}_i \cdot {\bf p}_j\right\rangle \; ,
\end{equation}
and the field shift (FS), which depends on the variation of the electron density  inside the nuclear charge distribution.
Using the wave functions of~\cite{Jonetal:10a}, the latter contribution is estimated to about 0.2~MHz in the considered transitions and is therefore neglected in the present work.
The level specific mass shift difference between two fine structure components $J'$ and $J$ of a same $LS$ term can be obtained by measuring the transition IS from these states to a common $L'S'J'$ level.

\paragraph{Upper levels $3p$ $^4P^o$ and $3p$ $^4D^o$ :}

Cangiano et al.~\cite{Canetal:94a} found some $J$-dependency for the upper \mbox{$3p$ $^4P^o$} term SMS. They obtained 110(300)~MHz and 318(300)~MHz for the SMS differences $5/2-3/2$ measured relatively to $3s$~$^4P_{5/2}$ and $3s$~$^4P_{3/2}$, respectively. Holmes \cite{Hol:43a} predicted a compatible shift of 51(33)~MHz. Only the value of Cangiano~et al. obtained with respect to the level $3s$~$^4P_{5/2}$ overlaps with the experimental results of Jennerich et al.~\cite{Jenetal:06a} who found a negative difference of $-$32.0(32)~MHz (see Table \ref{SMS}). 
In the case of the \mbox{$3p$ $^4D^o$} term, Jennerich~et al. measured $-$14.7(2.5)~MHz for the $7/2-5/2$ levels SMS difference.

As suggested in section~\ref{sec:cross},  the results of Cangiano et al.~\cite{Canetal:94a} and Jennerich et al.~\cite{Jenetal:06a} are affected by a wrong assignment of the spectral lines, inducing an error $\delta\nu_0$ on the fine structure transitions center. Adopting the present interpretation of the observed spectra, the SMS values are revised from the sum of the IS and $\delta\nu_0$, with their uncertainties (see last column of Table~\ref{SMS}). The error estimation on the SMS values is likely optimistic and should ultimately be refined from a proper fit of the recorded spectra. It is however useful within the limits exposed in the beginning of section~\ref{sec:cross}.

We observe from the third column of Table~\ref{SMS} a remarkably small $J$- and $L$-dependency of the SMS for the odd $3p$~$^4L^o_J$ upper states, in agreement with limited relativistic \emph{ab initio} calculations that estimate a $J$-dependency of maximum 1~MHz.

As discussed in section \ref{sec:nRes}, we predict unresolved spectra, as observed, for $3s~^4$P$_{J} \rightarrow 3p~^4$P$^o_{J'}$, with $(J,J')=(1/2,3/2)$, $(3/2,3/2)$ and $(3/2,5/2)$ (see Figure~\ref{nRes1a}, \ref{nRes1b} and \ref{nRes2}), while Jennerich et al.~\cite{Jenetal:06a} explained the absence of expected structure by invoking strong line shape perturbations. Therefore, the shifts for these lines  could be more meaningful that originally thought. However, a  fit of the original spectra  for those transitions is needed to assure a more precise determination of their SMS.

\paragraph{Lower level $^4P$ :} Holmes measured a value of $-$240(68) MHz for the SMS difference $^4P_{5/2-3/2}$  using the two transitions sharing the common $3p~^4S^o_{3/2}$ level~\cite{Hol:43a}. For the same difference, Cangiano et al. measured $-$553(300)~MHz and $-$344(300)~MHz using $3p$~$^4P^o_{5/2}$ and $3p$~$^4P^o_{3/2}$, respectively~\cite{Canetal:94a}. As revealed by the last column of Table~\ref{SMS}, all the SMS values for the $3s~^4P_{5/2} \rightarrow 3p~^4L^o_{J}$ transitions lie in a window of $-$2746(3)~MHz. Assuming that the SMS value differs weakly between the fine structure levels $3p~^4L^o_{J}$, as discussed above, we obtain about $-$167~MHz for the level SMS difference $3s$~$^4P$($5/2-3/2$) and about $-$91~MHz for ($3/2-1/2$).

This relatively large $J$-dependency can be explained by the well known strong mixing between $1s^2\ 2s^2\ 2p^2\ 3s\ ^4P$ and $1s^2\ 2s\ 2p^4\ ^4P$~\cite{Jonetal:10a,Hibetal:91b}. Indeed, the inspection of the Breit-Pauli eigenvectors obtained  in relatively large correlation spaces reveals that the weight of the $1s^2\ 2s\ 2p^4\ ^4P$  component ($\approx$0.3) increases  by about 1\% from  $J=1/2$ to $J=3/2$ levels, and by 2\% from $J=3/2$ to $J=5/2$. These changes of eigenvector compositions are most likely reliable since they well reproduce the observed fine structure (within 2\%). We conclude that the $3s$ $^4P$ $J$-dependency of the wave functions  can be estimated neglecting the term mixing of different $LS$-symmetries and is largely dominated by the relativistic effects on the $3s\ ^4P$ correlation.

\begin{table}[h]
\begin{center}
\caption{Transition specific mass shifts (in MHz). Comparison between the  values deduced from Jennerich et al.'s analysis~\cite{Jenetal:06a} (second column) with the values deduced from the
$\mathcal{S}_c$ spectra (present analysis - see text). The field shift is neglected.  \label{SMS}}
\begin{tabular}{ccc}
\toprule
Transition & ref.~\cite{Jenetal:06a} & This work\\
\midrule
$3s$ $^4$P$_{1/2} \rightarrow 3p~^4$D$^o_{1/2}$& $-$2488.1(15) & $-$2488.1(15)\\
$3s$ $^4$P$_{3/2} \rightarrow 3p~^4$P$^o_{1/2}$& $-$2558.3(22) & $-$2579.4(68)\\
$3s$ $^4$P$_{5/2} \rightarrow 3p~^4$D$^o_{5/2}$& $-$2748.17(84) & $-$2748.17(84)\\
$3s$ $^4$P$_{5/2} \rightarrow 3p~^4$D$^o_{7/2}$& $-$2762.9(16) & $-$2746.4(18) \\
$3s$ $^4$P$_{5/2} \rightarrow 3p~^4$P$^o_{3/2}$& $-$2713.4(14) & $-$2746.4(17)\\
$3s$ $^4$P$_{5/2} \rightarrow 3p~^4$P$^o_{5/2}$& $-$2745.4(18) & $-$2745.4(18)\\
\bottomrule
\end{tabular}
\end{center}
\end{table}

\section{Conclusion}


We completely revisited the analysis of the near-infrared hyperfine Nitrogen spectra for transitions $2p^{2}(^{3}$P$)$ $3s~^{4}$P$_{J}$ $ \rightarrow 2p^{2}(^{3}$P$)$ $3p~^{4}$P$^{o}_{J}$ and $^{4}$D$^{o}_{J}$. The proposed assignments for most of the weak lines observed by Cangiano et al.~\cite{Canetal:94a} and Jennerich et al.~\cite{Jenetal:06a} are built on the hypothesis of crossovers signals appearing 
with intensities comparable to the expected (weak) real transitions, while the latter do not appear in the experimental spectra. This suggests  strong perturbations in the recorded spectra, making the signals corresponding to weak transitions less intense than expected. 

The possibility of an improper assignment of hyperfine components was ruled out
by Jennerich~et al.~\cite{Jenetal:06a} by the fact that ``\emph{the fits shown in Figure~2 are so good\footnote{see top spectra of Figures~\ref{fig:Per_5/2_3/2} to \ref{NDP2} of the present paper.}, and the resulting  transition strength ratios are very close to the theoretical values}". On the other hand, for the transitions  $\; ^{4}$P$_{1/2}\rightarrow\; ^{4}$P$^{o}_{3/2}$ and $\; ^{4}$P$_{3/2}\rightarrow\; ^{4}$P$^{o}_{5/2}$, the original analysis~\cite{Jenetal:06a} called for strong line shape perturbations for explaining the non-observation of the expected resolved hyperfine structures for some transitions (one
or two of the hyperfine components of a given transition becoming dominant, and the others becoming negligible in strength).
The robustness of the present interpretation of the hyperfine spectra lies in the very good agreement of the present model with the observed non-resolved spectra for the latter transitions. Moreover, while systematic large theory-observation discrepancies appeared for the relevant hyperfine parameters~\cite{Jonetal:10a}, the present analysis provides an experimental estimation  (from the same spectra) of the hyperfine parameters in very good agreement with the {\em ab initio} results.

Non-linearities in the line intensities ratios are to be expected in saturated absorption spectroscopy. Even if the experimental setup can often be adapted to permit an unambiguous assignment of the spectra, we showed that this ambiguity can persist or, worse, that the spectra can be misleading, even, and maybe more, in very simple spectra. In those situations, theoretical calculations are  helpful in discriminating two probable scenarios.

The recorded spectra of Jennerich~et al.~\cite{Jenetal:06a} should be reinvestigated according to the present analysis to refine the new set of hyperfine constants set and associated uncertainties. A definitive confirmation of one set or another would be the observation of a signal that is predicted in one model, but not in the other. An alternative would be to show a crossover-like dependence of the weak lines intensities with the experimental setup.

Isotope shift values were extracted from Jennerich~et al.'s  spectra~\cite{Jenetal:06a}.  Significant variations of the IS within each multiplet were reported. The present analysis built on a substantial revision of the hyperfine line assignments washes out the $J$-dependency of SMS found for $3p$~$^4P^o$ and $3p$~$^4D^o$ multiplets. On the contrary, a somewhat large SMS $J$-dependency is deduced for the even parity $3s$~$^4P$  multiplet. This effect is enhanced by the strong non-relativistic mixing with $1s^2\ 2s\ 2p^4\ ^4P$, which depends strongly of the total atomic electronic momentum $J$ once relativistic corrections are added.

\begin{acknowledgement}

 TC is grateful to the ``Fonds pour la formation à la Recherche dans l'Industrie et dans l'Agriculture'' of Belgium for a PhD grant (Boursier F.R.S.-FNRS). MG thanks the Communaut\'e fran\c{c}aise of Belgium (Action de Recherche Concert\'ee)
and the Belgian National Fund for Scientific Research (FRFC/IISN Convention) for financial support. PJ acknowledges financial support from the Swedish Research Council.

\end{acknowledgement}


\end{document}